\documentclass[11pt]{article}
\usepackage{amssymb,amsmath,amsfonts}
\usepackage{graphicx}
\usepackage{graphics}
\usepackage{eepic,epsfig}

\@addtoreset{equation}{section}

\makeatother

\begin{document}

\title{Fermionic vacuum polarization around a cosmic string in compactified
AdS spacetime}
\author{S. Bellucci$^{1}$\thanks{%
E-mail: bellucci@lnf.infn.it },\, W. Oliveira dos Santos$^{1,2}$\thanks{%
E-mail: wagner.physics@gmail.com},\, E. R. Bezerra de Mello$^{2}$\thanks{%
E-mail: emello@fisica.ufpb.br},\, A. A. Saharian$^{3}$\thanks{%
E-mail: saharian@ysu.am} \\
\\
\textit{$^1$ INFN, Laboratori Nazionali di Frascati,}\\
\textit{Via Enrico Fermi 54, 00044 Frascati, Italy} \vspace{0.3cm}\\
\textit{$^{2}$Departamento de F\'{\i}sica, Universidade Federal da Para\'{\i}%
ba}\\
\textit{58.059-970, Caixa Postal 5.008, Jo\~{a}o Pessoa, PB, Brazil}\vspace{%
0.3cm}\\
\textit{$^3$Department of Physics, Yerevan State University,}\\
\textit{1 Alex Manoogian Street, 0025 Yerevan, Armenia}}
\maketitle

\begin{abstract}
We investigate topological effects of a cosmic string and compactification
of a spatial dimension on the vacuum expectation value (VEV) of the
energy-momentum tensor for a fermionic field in (4+1)-dimensional locally
AdS spacetime. The contribution induced by the compactification is
explicitly extracted by using the Abel-Plana summation formula. The mean
energy-momentum tensor is diagonal and the vacuum stresses along the
direction perpendicular to the AdS boundary and along the cosmic string are
equal to the energy density. All the components are even periodic functions
of the magnetic fluxes inside the string core and enclosed by compact
dimension, with the period equal to the flux quantum. The vacuum energy
density can be either positive or negative, depending on the values of the
parameters and the distance from the string. The topological contributions
in the VEV of the energy-momentum tensor vanish on the AdS boundary. Near
the string the effects of compactification and gravitational field are weak
and the leading term in the asymptotic expansion coincides with the
corresponding VEV in (4+1)-dimensional Minkowski spacetime. At large
distances, the decay of the cosmic string induced contribution in the vacuum
energy-momentum tensor, as a function of the proper distance from the
string, follows a power law. For a cosmic string in the Minkowski bulk and
for massive fields the corresponding fall off is exponential. Within the
framework of the AdS/CFT correspondence, the geometry for conformal field
theory on the AdS boundary corresponds to the standard cosmic string in
(3+1)-dimensional Minkowski spacetime compactified along its axis.
\end{abstract}


\bigskip

\section{Introduction}

\label{Int}

The formation of topological defects is among the interesting implications
of symmetry breaking phase transitions \cite{Hind95,Vile00}. This phenomenon
plays an important role in cosmology, in astrophysics and in a wide range of
condensed-matter systems. Within the framework of unified gauge theories of
particle interactions a sequence of phase transitions, accompanied by the
formation of a variety of topologically stable structures, is predicted in
the early Universe. Depending on the type of symmetry that is broken, the
corresponding defects can be cosmic strings, domain walls, monopoles and
textures. The investigation of physical effects induced by those defects is
an important issue to understand the early Universe. The corresponding
observable consequences provide a link between high-energy particle physics
and cosmology. Among other topological defects, the astrophysical and
cosmological implications of cosmic strings are most thoroughly studied. The
early interest was motivated by the possibility for cosmic strings to seed
perturbations into the energy density of the early Universe that can serve
as sources for the formation of the large-scale structures \cite{Durr02}.
The spectrum of primordial density perturbations is encoded in statistical
characteristics of the cosmic microwave background (CMB) radiation and
recent precision measurements of the CMB temperature anisotropy have
excluded the cosmic strings as the main source for perturbations. However,
the cosmic strings may have effects on the CMB properties that include the
creation of small non-Gaussianities and influence on the tensor modes. Among
the other interesting effects sourced by cosmic strings we mention here the
gravitational lensing, the radiation of gravitational waves, and the
generation of gamma ray bursts and high-energy cosmic rays. As it has been
discussed recently in \cite{Bran21}, cosmic string loops may seed black
holes with a continuous range of masses. This mechanism is especially
interesting in the case of intermediate mass black holes. An alternative
mechanism for the formation of cosmic string-type linear structures has been
suggested in brane inflation model within the framework of fundamental
string theory (for reviews see \cite{Sake07}-\cite{Cher15}).

The tension of cosmic string is proportional to the square of the symmetry
breaking energy scale and the corresponding energy density is localized
inside the core with the radius determined by the Compton wavelength of the
fields originating the cosmic string (complex scalar and gauge fields in
Abelian Higgs theory). Away from the core the geometry around a straight
cosmic string is well approximated by conical spacetime with planar angle
deficit determined by the string tension. Despite the fact that this
geometry is flat, the corresponding nontrivial topology gives rise to a
number of interesting physical effects. In particular, the vacuum
fluctuations of quantum fields are modified and, as a consequence, the
vacuum expectation values (VEVs) of local physical characteristics receive
topological contributions. Among those characteristics the VEV of the
energy-momentum tensor is of special interest. In addition to describing the
distribution of the energy density and vacuum stresses around the cosmic
string, it appears as a source of gravity in the semiclassical Einstein
equations and determines the backreaction of quantum effects on the
gravitational field. The VEV of the energy-momentum tensors for the
electromagnetic field and for massless scalar and fermionic fields in the
idealized geometry of a straight cosmic string with zero thickness core has
been investigated in \cite{Hell86}-\cite{Dowk87b}. Massive scalar and
fermionic fields have been discussed in \cite{Shir92}-\cite{Site12}. The
presence of boundaries induces additional contributions in the VEV\ of the
energy-momentum tensor (the Casimir effect). The combined effects of
topology and of various types of boundaries in the geometry of a cosmic
string have been discussed in \cite{Brev95}-\cite{Beze18perp}. The
compactification of cosmic string along its axis gives arise to topological
Casimir contributions in the vacuum energy-momentum tensor \cite{Beze12top}-%
\cite{Brag19}. For cosmic strings in curved backgrounds additional vacuum
polarization is induced by gravitational fields. Quantum effects near cosmic
strings in the Schwarzschild geometry have been studied in \cite{Davi88}-%
\cite{Otte11}. The VEVs of the energy momentum tensor for scalar, fermionic
and electromagnetic fields around cosmic strings in de Sitter and anti-de
Sitter (AdS) spacetimes were investigated in \cite{Beze09dS}-\cite%
{Oliv20compAdS}. It has been shown that the gravitational field essentially
changes the behavior of the vacuum densities at distances from the string
larger than the curvature radius of the background spacetime.

Continuing our previous study of the combined effects of cosmic string,
background gravitational field and compactification on the local properties
of the vacuum state, in the present paper we investigate the vacuum
energy-momentum tensor for a massive fermionic field in background of
(4+1)-dimensional locally AdS spacetime with a compactified dimension and in
the presence of a cosmic string. The VEV of the current density and the
fermion condensate in the same geometry have been considered recently in 
\cite{Bell20,Bell21FC}. The effects induced by a brane parallel to the AdS
boundary were discussed in \cite{Bell21AdSb}. Our choice of the AdS
spacetime as a background geometry is motivated by its crucial role in
braneworld models with extra dimensions and in AdS/CFT correspondence (for
reviews see \cite{Maar10,Ahar00,Ammo15}). It is also important that the AdS
spacetime is maximally symmetric and closed analytic expressions can be
obtained for physical characteristics of the vacuum state. The
corresponding results may shed light on the influence of gravitational
fields on quantum matter in more complicated geometries.

The organization of the paper is as follows. The problem setup is presented
in the next section. The VEV of the energy-momentum tensor in the geometry
of a straight cosmic string in background of AdS spacetime is studied in
Section \ref{sec:EMTcs}. The contribution in the vacuum energy-momentum
tensor induced by compactification of a spatial dimension is considered in
Section \ref{sec:EMTcomp}. The main results of the paper are summarized in
Section \ref{sec:Conc}. In Appendix \ref{sec:App1}, an alternative
representation for the topological contribution in the VEV of the
energy-momentum tensor is provided.

\section{Problem setup}

\label{sec:Setup}

In the presence of an external gauge field with the vector potential $A_{\mu
}$ and on a spacetime background described by the metric tensor $g_{\mu \nu
} $, the quantum fermionic field $\psi (x)$ obeys the equation 
\begin{equation}
\left( i\gamma ^{\mu }\mathcal{D}_{\mu }-sm\right) \psi (x)=0\,,\;\mathcal{D}%
_{\mu }=\partial _{\mu }+\Gamma _{\mu }+ieA_{\mu },  \label{DiracEq}
\end{equation}%
where the curved spacetime Dirac matrices $\gamma ^{\mu }$ form the Clifford
algebra $\left\{ \gamma ^{\mu },\gamma ^{\nu }\right\} =2g^{\mu \nu }$ and $%
\Gamma _{\mu }$ is the spin connection. The matrices $\gamma ^{\mu }$ are
related to the corresponding flat spacetime Dirac matrices $\gamma ^{(a)}$
through the vielbein fields $e_{(a)}^{\mu }$: $\gamma ^{\mu }=e_{(a)}^{\mu
}\gamma ^{(a)}$. The spin connection is expressed in terms of those fields
as $\Gamma _{\mu }=\gamma ^{(a)}\gamma ^{(b)}e_{(a)}^{\nu }\nabla _{\mu
}e_{(b)\nu }/4$, where $\nabla _{\mu }e_{(b)\nu }$ stands for the covariant
derivative of the vector field with the covariant components determined by
the index $\nu $. The parameter $s$ takes the values $s=+1$ and $s=-1$ and
corresponds to the two irreducible representations of the Clifford algebra
in odd-dimensional spacetimes (this point will be discussed in detail below).

The background geometry under consideration in the present paper is
described by the (4+1)-dimensional line element 
\begin{equation}
ds^{2}=e^{-2y/a}\left( dt^{2}-dr^{2}-r^{2}d\phi ^{2}-dz^{2}\right) -dy^{2}\ ,
\label{ds1}
\end{equation}%
with the spacetime coordinates $x^{\mu }=(t,r,\phi ,y,z)$ varying in the
ranges $-\infty <t,y<+\infty $, $0\leq r<\infty $, and $0\leq \phi \leq \phi
_{0}$. Introducing a new coordinate $w=ae^{y/a}$, $0\leq w<\infty $, the
line element is presented in the conformally-flat form (for $r>0$) with the
metric tensor $g_{\mu \nu }=(a/w)^{2}\mathrm{diag}(1,-1,-r^{2},-1,-1)$. For
the $z$-coordinate we will consider two cases. In the first one $-\infty
<z<+\infty $ and for the special value $\phi _{0}=2\pi $, the line element (%
\ref{ds1}) corresponds to the slice of the AdS spacetime described in Poincar%
\'{e} coordinates. For the corresponding cosmological constant one has $%
\Lambda =-6/a^{2}$ and the Ricci tensor is expressed as $R_{\mu \nu
}=-4g_{\mu \nu }/a^{2}$. For $\phi _{0}<2\pi $ the line element (\ref{ds1})
generalizes the idealized geometry of a cosmic string for (4+1)-dimensional
background AdS spacetime. Although the surface $r=0$ corresponds to a
two-dimensional spatial surface with the induced line element $%
dy^{2}+e^{-2y/a}dz^{2}$, in the discussion below we will use the term cosmic
string. The (3+1)-dimensional geometry of a cosmic string on the AdS bulk
(see \cite{Dehg02,Ball11,Padu16}) corresponds to the hypersurface $z=\mathrm{%
const}$. The geometry of another hypersurface with $y=\mathrm{const}$
corresponds to the standard geometry of a cosmic string in (3+1)-dimensional
Minkowski spacetime. In particular, this also applies to the background
geometry for the conformal field theory within the framework of the AdS/CFT
correspondence. For $r>0$ the local geometry corresponding to (\ref{ds1})
coincides with that for the AdS spacetime. The second case corresponds to
the $z$-direction compactified on a circle and it will be discussed in
Section \ref{sec:EMTcomp}.

We are interested in the effects of the topology change, induced by the
cosmic string and by compactification, on the VEV of the energy-momentum
tensor $\left\langle 0|T_{\mu \nu }|0\right\rangle =\left\langle T_{\mu \nu
}\right\rangle $ for the fermionic field $\psi (x)$. For the gauge field, a
simple configuration $A_{\mu }=(0,0,A_{2},0,A_{4})$ will be considered,
where the covariant components $A_{2}$ and $A_{4}$ are constants. The
component $A_{2}$ is expressed in terms of the magnetic flux $\Phi _{\mathrm{%
s}}$ running along the string's core as $A_{2}=-q\Phi _{\mathrm{s}}/(2\pi )$%
. In the problem with uncompactified $z$-direction the component $A_{4}$
will not appear in the expressions for the VEVs. That component is
physically relevant for the geometry with compact $z$-coordinate.

The VEV of the energy-momentum tensor can be evaluated by using the mode-sum
formula 
\begin{equation}
\left\langle T_{\mu \nu }\right\rangle =-\frac{i}{4}\sum_{\sigma }\sum_{\chi
=-,+}\chi {\left[ \bar{\psi}_{\sigma }^{(\chi )}\gamma _{(\mu }\mathcal{D}%
_{\nu )}\psi _{\sigma }^{(\chi )}-(\mathcal{D}_{(\mu }\bar{\psi}_{\sigma
}^{(\chi )})\gamma _{\nu )}\psi _{\sigma }^{(\chi )}\right] }\ ,  \label{EMT}
\end{equation}%
where $\{\psi _{\sigma }^{(+)},\psi _{\sigma }^{(-)}\}$ is the complete set
of the positive and negative energy fermionic modes, specified by the set of
quantum numbers $\sigma $. In (\ref{EMT}), the brackets in the index
expression mean the symmetrization over the enclosed indices, the Dirac
adjoint is defined as $\bar{\psi}_{\sigma }^{(\chi )}=\psi _{\sigma }^{(\chi
)\dagger }\gamma ^{(0)}$ and $\mathcal{D}_{\mu }{\bar{\psi}}_{\sigma
}^{(\chi )}=\partial _{\mu }{\bar{\psi}}_{\sigma }^{(\chi )}-ieA_{\mu }{\bar{%
\psi}}_{\sigma }^{(\chi )}-{\bar{\psi}}_{\sigma }^{(\chi )}\Gamma _{\mu }$
for $\chi =+,-$. The expression in the right-hand side of (\ref{EMT}) is
divergent and a regularization is required. For example, we can introduce a
cutoff function or employ the point-splitting procedure. In the discussion
below we will extract from the VEV the topological part. For $r>0$ it is
finite and does not depend on the specific regularization procedure.

In what follows it will be convenient to work in the coordinate system $%
x^{\mu }=(t,r,\phi ,w,z)$ with the conformal coordinate $w$. We will take
the vielbein fields $e_{(b)}^{\mu }=\delta _{b}^{\mu }w/a$ and the flat
spacetime Dirac matrices 
\begin{eqnarray}
\gamma ^{(0)} &=&\left( {%
\begin{array}{cc}
0 & -i \\ 
i & 0%
\end{array}%
}\right) \ ,\;\gamma ^{(4)}=\left( {%
\begin{array}{cc}
0 & i \\ 
i & 0%
\end{array}%
}\right) ,  \notag \\
\gamma ^{(l)} &=&-i\,\mathrm{diag}(\sigma ^{l},-\sigma ^{l}),\;l=1,2,3.
\label{Dirac04}
\end{eqnarray}%
Here, $2\times 2$ Pauli matrices in the cylindrical coordinates $(r,\phi ,w)$
are given by 
\begin{eqnarray}
\sigma ^{l} &=&\left( \frac{i}{r}\right) ^{l-1}\left( {%
\begin{array}{cc}
0 & (-1)^{l-1}e^{-iq\phi } \\ 
e^{iq\phi } & 0%
\end{array}%
}\right) \ ,  \notag \\
\sigma ^{3} &=&\mathrm{diag}(1,-1),\;l=1,2,  \label{PauliCyl}
\end{eqnarray}%
with the notation $q=2\pi /\phi _{0}$. The geometries with uncompact and
compact $z$-directions will be considered separately.

\section{Vacuum energy-momentum tensor in the uncompactified \newline
geometry}

\label{sec:EMTcs}

In this section we consider the VEV of the energy-momentum tensor in the
geometry with uncompactified $z$-direction, $-\infty <z<+\infty $. The
corresponding fermionic modes are specified by the set of quantum numbers $%
\sigma =(\lambda ,p,k_{z},j,\eta )$ with the variation ranges $0\leq \lambda
,p<\ \infty $, $-\infty <k_{z}<+\infty $, $j=\pm 1/2,\pm 3/2,...$, $\eta
=\pm 1$. The mode functions are expressed as (for the modes with compact $z$%
-dimension see \cite{Bell20}) 
\begin{equation}
\psi _{\sigma }^{(\pm )}(x)=C_{\sigma }^{(\pm )}w^{5/2}\left( {%
\begin{array}{c}
J_{\beta _{j}}(\lambda r)J_{\nu _{1}}(pw)e^{-iq\phi /2} \\ 
\mp s\epsilon _{j}\kappa _{\eta }b_{\eta }^{(\pm )}J_{\beta _{j}+\epsilon
_{j}}(\lambda r)J_{\nu _{2}}(pw)e^{iq\phi /2} \\ 
is\kappa _{\eta }J_{\beta _{j}}(\lambda r)J_{\nu _{2}}(pw)e^{-iq\phi /2} \\ 
\pm i\epsilon _{j}b_{\eta }^{(\pm )}J_{\beta _{j}+\epsilon _{j}}(\lambda
r)J_{\nu _{1}}(pw)e^{iq\phi /2}%
\end{array}%
}\right) e^{iqj\phi +ik_{z}z\mp iEt}\ ,  \label{Modes1}
\end{equation}%
where $\nu _{l}=ma+(-1)^{l}s/2$, $J_{\nu }(x)$ is the Bessel function \cite%
{Abra,Nist10} and the energy is given by $E=\sqrt{\lambda
^{2}+p^{2}+k_{z}^{2}}$. The order of the Bessel function is defined by 
\begin{equation}
\beta _{j}=q|j+\alpha |-\epsilon _{j}/2\ ,  \label{order_Bessel}
\end{equation}%
where $\epsilon _{j}=\mathrm{sgn}(j+\alpha )$ with $\alpha =eA_{2}/q=-\Phi _{%
\mathrm{s}}/\Phi _{0}$. Here $\Phi _{\mathrm{s}}$ is the magnetic flux
running along the string's core and $\Phi _{0}$ is the flux quantum. The
coefficients $\kappa _{\eta }$ and $b_{\eta }^{(\pm )}$ are given by the
expressions 
\begin{eqnarray}
&&\kappa _{\eta }=\frac{1}{p}\left( -k_{z}+\eta \sqrt{p^{2}+k_{z}^{2}}%
\right) ,  \notag \\
&&b_{\eta }^{(\pm )}=\frac{1}{\lambda }\left( E\mp \eta \sqrt{p^{2}+k_{z}^{2}%
}\right) \,.  \label{kb}
\end{eqnarray}%
For the normalization coefficient one gets 
\begin{equation}
|C_{\sigma }^{(\pm )}|^{2}=\frac{\left( 4\pi \right) ^{-2}\eta
a^{-4}qp^{2}\lambda ^{2}}{E\kappa _{\eta }b_{\eta }^{(\pm )}\sqrt{%
p^{2}+k_{z}^{2}}}\ .  \label{coeff}
\end{equation}

Having the normalized mode functions (\ref{Modes1}), the VEV of the
energy-momentum tensor is evaluated by using the formula (\ref{EMT}), where
the summation over $\sigma $ in is understood as 
\begin{equation}
\sum_{\sigma }=\sum_{j}\int_{0}^{\infty }d\lambda \int_{0}^{\infty
}dp\int_{-\infty }^{\infty }dk_{z}\sum_{\eta =\pm 1}\,,  \label{sumj}
\end{equation}%
with $\sum_{j}=\sum_{j=\pm 1/2,\pm 3/2,\cdots }$. We can show that the VEVs
for $s=+1$ and $s=-1$ coincide and in what follows the formulas will be
presented for $s=+1$. The VEVs of the diagonal components are presented as
(no summation over $\mu $)%
\begin{equation}
\left\langle T_{\mu }^{\mu }\right\rangle _{\mathrm{cs}}^{\mathrm{AdS}}=-%
\frac{qw^{6}}{4\pi ^{2}a^{5}}\sum_{j}\int_{0}^{\infty }d\lambda \lambda
\int_{0}^{\infty }dp\,p\int_{0}^{\infty }dk_{z}\,E^{2\delta _{0\mu
}-1}\left( k_{z}^{2}\right) ^{\delta _{4\mu }}R_{\beta _{j}}^{(\mu )}\left(
\lambda r\right) W_{\nu }^{(\mu )}(pw),  \label{Tkk}
\end{equation}%
where $\nu =ma-1/2$,%
\begin{eqnarray}
R_{\beta _{j}}^{(0)}\left( \lambda r\right) &=&R_{\beta _{j}}^{(3)}\left(
\lambda r\right) =-R_{\beta _{j}}^{(4)}\left( \lambda r\right) =J_{\beta
_{j}}^{2}(\lambda r)+J_{\beta _{j}+\epsilon _{j}}^{2}(\lambda r),  \notag \\
R_{\beta _{j}}^{(1)}\left( \lambda r\right) &=&\epsilon _{j}\lambda ^{2} 
\left[ J_{\beta _{j}}^{\prime }(\lambda r)J_{\beta _{j}+\epsilon
_{j}}(\lambda r)-J_{\beta _{j}}(\lambda r)J_{\beta _{j}+\epsilon
_{j}}^{\prime }(\lambda r)\right] ,  \notag \\
R_{\beta _{j}}^{(2)}\left( \lambda r\right) &=&-\frac{\lambda }{r}(2\beta
_{j}+\epsilon _{j})J_{\beta _{j}}(\lambda r)J_{\beta _{j}+\epsilon
_{j}}(\lambda r),  \label{R2}
\end{eqnarray}%
and%
\begin{eqnarray}
W_{\nu }^{(\mu )}(pw) &=&J_{\nu }^{2}(pw)+J_{\nu +1}^{2}(pw),\;\mu =0,1,2,4,
\notag \\
W_{\nu }^{(3)}(pw) &=&p^{2}\left[ J_{\nu }^{\prime }(pw)J_{\nu
+1}(pw)-J_{\nu }(pw)J_{\nu +1}^{\prime }(pw)\right] .  \label{W3}
\end{eqnarray}%
Note that we have the relation%
\begin{equation}
R_{\beta _{j}}^{(1)}\left( \lambda r\right) =\lambda ^{2}R_{\beta
_{j}}^{(0)}\left( \lambda r\right) +R_{\beta _{j}}^{(2)}\left( \lambda
r\right) .  \label{R1}
\end{equation}%
The off-diagonal components of the vacuum energy-momentum tensor vanish.

For the extraction of the topological part in (\ref{Tkk}) it is convenient
to use the integral representation 
\begin{equation}
E^{2\delta _{0\mu }-1}=\frac{2}{\sqrt{\pi }}\int_{0}^{\infty }d\tau \,\left(
\partial _{-\tau ^{2}}\right) ^{\delta _{0\mu }}e^{-(\lambda
^{2}+p^{2}+k_{z}^{2})\tau ^{2}}.  \label{Id}
\end{equation}%
After evaluating the integral over $k_{z}$, the $p$-integral is evaluated by
using the formula%
\begin{equation}
\int_{0}^{\infty }dppe^{-p^{2}\tau ^{2}}W_{\nu }^{(\mu )}(pw)=\frac{\left(
-1\right) ^{\delta _{3\mu }}e^{-x}}{(2\tau ^{2})^{1+\delta _{3\mu }}}\left[
I_{\nu }(x)+I_{\nu +1}(x)\right] _{x=w^{2}/2\tau ^{2}},  \label{Intp}
\end{equation}%
where $I_{\nu }(x)$ is the modified Bessel function \cite{Abra}. The
integral over $\lambda $ involving the functions $R_{\beta _{j}}^{(\mu
)}\left( \lambda r\right) $ with $\mu \neq 2$ has the structure similar to (%
\ref{Intp}) and the integral for $R_{\beta _{j}}^{(2)}\left( \lambda
r\right) $ is evaluated by using the relation (\ref{R1}). The components
with $\mu =0,1,3,4$ are presented in the form (no summation over $\mu $)%
\begin{equation}
\left\langle T_{\mu }^{\mu }\right\rangle _{\mathrm{cs}}^{\mathrm{AdS}}=%
\frac{qa^{-5}}{8\pi ^{2}}\int_{0}^{\infty }dx\,x^{2}e^{-x(1+\rho ^{2})}\left[
I_{\nu }(x)+I_{\nu +1}(x)\right] {\mathcal{J}}(q,\alpha _{0},x\rho ^{2}),
\label{Tkk2}
\end{equation}%
with $\rho =r/w$ and 
\begin{equation}
{\mathcal{J}}(q,\alpha ,y)=\sum_{j}\left[ I_{\beta _{j}}\left( y\right)
+I_{\beta _{j}+\epsilon _{j}}\left( y\right) \right] .  \label{Jcal}
\end{equation}%
It can be seen that the remaining component is related to the energy density
by the formula%
\begin{equation}
\langle T_{2}^{2}\rangle _{\mathrm{cs}}^{\mathrm{AdS}}=\left( 1+r\partial
_{r}\right) \langle T_{0}^{0}\rangle _{\mathrm{cs}}^{\mathrm{AdS}}\ .
\label{T22}
\end{equation}%
The proper distance from the string is given by 
\begin{equation}
r_{p}=ar/w  \label{rp}
\end{equation}%
and $\rho $ in (\ref{Tkk2}) is the proper distance measured in units of the
curvature radius $a$. Note that the parameter $\alpha $ enters in the
expression for the VEV\ $\left\langle T_{\mu }^{\mu }\right\rangle _{\mathrm{%
cs}}^{\mathrm{AdS}}$ in the form $j+\alpha $. Consequently, if we present it
as $\alpha =\alpha _{0}+n_{0}$, with $|\alpha _{0}|<1/2$, then, after the
redefinition $j+n_{0}\rightarrow j$, we see that the VEV does not depend on
the integer part $n_{0}$. Hence, in the discussion below we can take $\alpha
=\alpha _{0}$ without loss of generality.

For the further transformation of the VEV we use the representation \cite%
{Beze10f3}%
\begin{eqnarray}
{\mathcal{J}}(q,\alpha _{0},y) &=&\frac{2}{q}e^{y}+\frac{4}{q}%
\sideset{}{'}{\sum}_{k=1}^{[q/2]}(-1)^{k}c_{k}\cos (2\pi k\alpha
_{0})e^{y\cos (2\pi k/q)}  \notag \\
&&+\frac{4}{\pi }\int_{0}^{\infty }du\frac{H(q,\alpha _{0},u)e^{-y\cosh 2u}}{%
\cosh (2qu)-\cos (q\pi )},  \label{Jf}
\end{eqnarray}%
where $[q/2]$ is the integer part of $q/2$ and the prime on the summation
sign means that for even values of $q$ the term with $k=q/2$ should be
halved. In (\ref{Jf}), the notations $c_{k}=\cos (\pi k/q)$ and 
\begin{equation}
H(q,\alpha _{0},u)=\sinh u\sum_{\chi =+,-}\cos [\pi q(1/2+\chi \alpha
_{0})]\sinh [(1-2\chi \alpha _{0})qu]  \label{hq}
\end{equation}%
have been introduced. In the special case $\alpha _{0}=0$ (the magnetic flux
along the string is a multiple of the flux quantum) the expression (\ref{hq}%
) is simplified to%
\begin{equation}
H(q,0,u)=2\cos (\pi q/2)\sinh (qu)\sinh u,  \label{Hq0}
\end{equation}%
and in (\ref{Jf}) the integral term vanishes for odd values of $q$.

Substituting the representation (\ref{Jf}) in (\ref{Tkk2}), the VEV is
decomposed as (no summation over $\mu $) 
\begin{equation}
\left\langle T_{\mu }^{\mu }\right\rangle _{\mathrm{cs}}^{\mathrm{AdS}%
}=\left\langle T_{\mu }^{\mu }\right\rangle ^{\mathrm{AdS}}+\left\langle
T_{\mu }^{\mu }\right\rangle _{\mathrm{cs}}.  \label{Tmu1}
\end{equation}%
The first contribution in the right-hand side comes from the first term in
the right-hand side of (\ref{Jf}) and is given by the expression 
\begin{equation}
\left\langle T_{\mu }^{\mu }\right\rangle ^{\mathrm{AdS}}=\frac{a^{-5}}{4\pi
^{2}}\int_{0}^{\infty }dx\,x^{2}e^{-x}\left[ I_{\nu }(x)+I_{\nu +1}(x)\right]
,  \label{TkkAdS}
\end{equation}%
for all values of $\mu $. The part $\left\langle T_{\mu }^{\mu
}\right\rangle _{\mathrm{cs}}$ in (\ref{Tmu1}) comes from the second and
third terms in the right-hand side of (\ref{Jf}) and for the components with 
$\mu =0,1,3,4$ it is presented as 
\begin{equation}
\left\langle T_{\mu }^{\mu }\right\rangle _{\mathrm{cs}}=\frac{a^{-5}}{\sqrt{%
2}\pi ^{5/2}}\left[ \sideset{}{'}{\sum}_{k=1}^{[q/2]}(-1)^{k}c_{k}\cos (2\pi
k\alpha _{0}){\mathcal{F}}_{ma}(u_{k})+\frac{q}{\pi }\int_{0}^{\infty }dx%
\frac{H(q,\alpha _{0},x){\mathcal{F}}_{ma}(u_{x})}{\cosh (2qx)-\cos (q\pi )}%
\right] \ ,  \label{Tllcs}
\end{equation}%
with the notations 
\begin{eqnarray}
u_{k} &=&1+2\rho ^{2}s_{k}^{2},\;s_{k}=\sin (\pi k/q)\ ,  \notag \\
u_{x} &=&1+2\rho ^{2}\cosh ^{2}x\ ,  \label{args-cs}
\end{eqnarray}%
and%
\begin{equation}
{\mathcal{F}}_{ma}(u)=\sqrt{\frac{\pi }{2}}\int_{0}^{\infty }dx\,x^{2}e^{-ux}%
\left[ I_{\nu }(x)+I_{\nu +1}(x)\right] \,.  \label{Ffunc}
\end{equation}%
The component $\left\langle T_{2}^{2}\right\rangle _{\mathrm{cs}}$ is
obtained from 
\begin{equation}
\langle T_{2}^{2}\rangle _{\mathrm{cs}}=\left( 1+r\partial _{r}\right)
\langle T_{0}^{0}\rangle _{\mathrm{cs}}.  \label{T22cs}
\end{equation}%
The VEV $\left\langle T_{\mu }^{\mu }\right\rangle _{\mathrm{cs}}$ depends
on the coordinates $r$ and $w$ through the ratio $r/w$. This property is a
consequence of the maximal symmetry of the AdS spacetime. Note that the VEV $%
\left\langle T_{\mu }^{\mu }\right\rangle _{\mathrm{cs}}$ is an even
function of the parameter $\alpha _{0}$. In terms of the magnetic flux along
the string core, $\left\langle T_{\mu }^{\mu }\right\rangle _{\mathrm{cs}}$
is an even periodic function of the magnetic flux with the period of the
flux quantum.

We have shown that the mean energy-momentum tensor is diagonal. This
property is not a consequence of the problem symmetry. In general, we could
have a nonzero component $\left\langle T_{4}^{1}\right\rangle _{\mathrm{cs}}$
allowed by the symmetry. As it has been discussed in \cite{Beze12AdS}, this
type of nonzero off-diagonal component is present for a massive scalar field
around a cosmic string in $(D+1)$-dimensional AdS spacetime. Note that for a
scalar field, in general, the components $\left\langle
T_{0}^{0}\right\rangle _{\mathrm{cs}}$ and $\left\langle
T_{1}^{1}\right\rangle _{\mathrm{cs}}$ are different as well. It can be
checked that the VEV $\left\langle T_{\mu \nu }\right\rangle _{\mathrm{cs}}$
obeys the covariant continuity equation $\nabla _{\mu }\left\langle T_{\nu
}^{\mu }\right\rangle _{\mathrm{cs}}=0$ \ and the trace relation $%
\left\langle T_{\mu }^{\mu }\right\rangle _{\mathrm{cs}}=sm\left\langle \bar{%
\psi}\psi \right\rangle _{\mathrm{cs}}$, with the fermion condensate $%
\left\langle \bar{\psi}\psi \right\rangle _{\mathrm{cs}}$ from \cite%
{Bell21FC}. In particular, for a massless field the tensor $\left\langle
T_{\mu \nu }\right\rangle _{\mathrm{cs}}$ is traceless. The trace anomaly is
contained in the part $\left\langle T_{\mu }^{\mu }\right\rangle ^{\mathrm{%
AdS}}$. The continuity equation is reduced to the relations%
\begin{equation}
\left\langle T_{2}^{2}\right\rangle _{\mathrm{cs}}=\partial
_{r}(r\left\langle T_{1}^{1}\right\rangle _{\mathrm{cs}}),\;\left\langle
T_{2}^{2}\right\rangle _{\mathrm{cs}}=\left( 1-w\partial _{w}\right)
\left\langle T_{3}^{3}\right\rangle _{\mathrm{cs}}.  \label{ContEq}
\end{equation}
These relations directly follow from (\ref{T22cs}) by taking into account
that $\left\langle T_{1}^{1}\right\rangle _{\mathrm{cs}}=\left\langle
T_{3}^{3}\right\rangle _{\mathrm{cs}}=\left\langle T_{0}^{0}\right\rangle _{%
\mathrm{cs}}$ and the property that the VEV depends on $r$ and $w$ through
the ratio $r/w$.

The special case with $q=1$ and $\alpha _{0}=0$ corresponds to the absence
of the cosmic string and magnetic flux and the corresponding geometry is
reduced to the AdS spacetime. In this special case one has $\left\langle
T_{\mu }^{\mu }\right\rangle _{\mathrm{cs}}=0$ and, hence, the part $%
\left\langle T_{\mu }^{\mu }\right\rangle ^{\mathrm{AdS}}$ is the VEV in AdS
spacetime. For it one has $\left\langle T_{\mu \nu }\right\rangle ^{\mathrm{%
AdS}}=\mathrm{const}\cdot g_{\mu \nu }$ and this property is a consequence
of the maximal symmetry of AdS spacetime. For $r>0$ the contribution $%
\left\langle T_{\mu }^{\mu }\right\rangle _{\mathrm{cs}}$ is finite and the
renormalization is required for the pure AdS part $\left\langle T_{\mu
}^{\mu }\right\rangle ^{\mathrm{AdS}}$ only. This feature is a consequence
of the fact that in the region $r>0$ the local geometrical characteristics
are the same as those in AdS spacetime and, hence, the divergences are the
same as well. The AdS part $\left\langle T_{\mu \nu }\right\rangle ^{\mathrm{%
AdS}}$ is widely considered in the literature and here we are interested in
the topological effects induced by the cosmic string and magnetic flux.

The integral in (\ref{Ffunc}) is expressed in terms of the associated
Legendre function $Q_{\nu }^{\mu }(z)$ \cite{Abra,Nist10} and one finds%
\begin{equation}
{\mathcal{F}}_{ma}(u)=\frac{Q_{ma-1}^{5/2}(u)+Q_{ma}^{5/2}(u)}{e^{5i\pi
/2}(u^{2}-1)^{5/4}}.  \label{Ffunc2}
\end{equation}%
For the evaluation of the derivative $\partial _{u}{\mathcal{F}}_{ma}(u)$ in
(\ref{T22cs}) one can use the relation 
\begin{equation}
\partial _{u}{\mathcal{F}}_{ma}(u)=-\frac{Q_{ma-1}^{7/2}(u)+Q_{ma}^{7/2}(u)}{%
e^{7\pi i/2}(u^{2}-1)^{7/4}}.  \label{dF}
\end{equation}%
For a massless field $\nu =-1/2$ and the modified Bessel functions in (\ref%
{Ffunc}) are expressed in terms of elementary functions. In this special
case one gets%
\begin{equation}
\ {\mathcal{F}}_{0}(u)=\frac{3\sqrt{\pi }}{4(u-1)^{5/2}}.  \label{Fm0}
\end{equation}%
For the discussion of the behavior of the VEV in the asymptotic regions of
the parameters we will need the asymptotics of the function ${\mathcal{F}}%
_{ma}(u)$ for large values of the argument and for $u\rightarrow 1$. By
using the asymptotic expression of the function $Q_{\nu -1/2}^{5/2}(u)$ for $%
u\rightarrow 1+$ (see, for example, \cite{Nist10}) we can see that 
\begin{equation}
{\mathcal{F}}_{ma}(u)\approx \frac{3\sqrt{\pi }}{4(u-1)^{5/2}}.  \label{Fas}
\end{equation}%
Note that the leading term given by (\ref{Fas}) coincides with (\ref{Fm0}).
For large argument the leading term in the corresponding asymptotic
expansion is given by 
\begin{equation}
{\mathcal{F}}_{ma}(u)\approx \frac{\sqrt{\pi }(2ma+3)(2ma+1)}{%
2^{ma+2}u^{ma+5/2}}.  \label{Fas2}
\end{equation}%
This result is obtained by using the asymptotic for the associated Legendre
function \cite{Nist10} or directly from the integral representation (\ref%
{Ffunc}).

Let us consider some special and limiting cases of the general formula (\ref%
{Tllcs}). For a massless field, by using (\ref{Fm0}), one finds (no
summation over $\mu $)%
\begin{equation}
\left\langle T_{\mu }^{\mu }\right\rangle _{\mathrm{cs}}=\frac{%
3h_{5}(q,\alpha _{0})}{32\pi ^{2}r_{p}^{5}}\ ,  \label{Tllm0}
\end{equation}%
for $\mu =0,1,3,4$ and $\langle T_{2}^{2}\rangle _{\mathrm{cs}}=-4\langle
T_{0}^{0}\rangle _{\mathrm{cs}}$. Here and in what follows the notation%
\begin{equation}
h_{n}(q,\alpha _{0})=\sideset{}{'}{\sum}_{k=1}^{[q/2]}(-1)^{k}\frac{c_{k}}{%
s_{k}^{n}}\cos (2\pi k\alpha _{0})+\frac{q}{\pi }\int_{0}^{\infty }dx\,\frac{%
H(q,\alpha _{0},x)\cosh ^{-n}x}{\cosh (2qx)-\cos (q\pi )}\ ,  \label{hn}
\end{equation}%
is introduced. In the special case $q=1$, corresponding to the absence of
planar angle deficit, the expression (\ref{hn}) is simplified to 
\begin{equation}
h_{n}(1,\alpha _{0})=\frac{\sin (\pi \alpha _{0})}{\pi }\int_{0}^{\infty
}dx\,\frac{\sinh (2\alpha _{0}x)}{\cosh ^{n+1}x}\sinh x\ .  \label{hn1}
\end{equation}

As seen from (\ref{ds1}), the limit $a\rightarrow \infty $ for fixed $y$
corresponds to the geometry of a cosmic string in background of
(4+1)-dimensional Minkowski spacetime. In this limit $ma\gg 1$ and the order
of the associated Legendre function in (\ref{Ffunc2}) is large. For the $w$%
-coordinate one has $w\approx a+y$ and in (\ref{Tllcs}) $u_{k},u_{x}%
\rightarrow 1+$. Consequently, we need the asymptotic of the function $%
Q_{\nu -1/2}^{5/2}(u)$ for $\nu \gg 1$ and $u-1\ll 1$. By using that
asymptotic from \cite{Nist10}, we can see that%
\begin{equation}
{\mathcal{F}}_{ma}(u)\approx 2m^{5}a^{5}f_{5/2}(ma\sqrt{2\left( u-1\right) }%
),  \label{Funi}
\end{equation}%
where $f_{\mu }(x)=K_{\mu }(x)/x^{\mu }$ and $K_{\mu }(x)$ is the modified
Bessel function. As a result of the limiting transition we get the VEV of
the energy-momentum tensor induced by a cosmic string in (4+1)-dimensional
Minkowski spacetime (no summation over $\mu =0,1,3,4$): 
\begin{eqnarray}
\left\langle T_{\mu }^{\mu }\right\rangle _{\mathrm{cs}}^{\mathrm{(M)}} &=&%
\frac{\sqrt{2}m^{5}}{\pi ^{5/2}}\left[ \sideset{}{'}{\sum}%
_{k=1}^{[q/2]}(-1)^{k}c_{k}\cos (2\pi k\alpha _{0})f_{5/2}(2mrs_{k})\right. 
\notag \\
&+&\left. \frac{q}{\pi }\int_{0}^{\infty }dx\frac{H(q,\alpha
_{0},x)f_{5/2}(2mr\cosh x)}{\cosh (2qx)-\cos (q\pi )}\right] \ .
\label{TllM}
\end{eqnarray}%
The component $\langle T_{2}^{2}\rangle _{\mathrm{cs}}^{\mathrm{(M)}}$ is
found from the relation similar to (\ref{T22cs}). For the function $%
f_{5/2}\left( x\right) $ in this expression one has%
\begin{equation}
f_{5/2}\left( x\right) =\sqrt{\frac{\pi }{2}}\frac{x^{2}+3x+3}{x^{5}}e^{-x}.
\label{f52}
\end{equation}%
In the case of a massless field the formula (\ref{TllM}) is reduced to%
\begin{equation}
\left\langle T_{\mu }^{\mu }\right\rangle _{\mathrm{cs}}^{\mathrm{(M)}}=%
\frac{3h_{5}(q,\alpha _{0})}{32\pi ^{2}r^{5}}.  \label{TllMm0}
\end{equation}%
This result is also seen from (\ref{Tllm0}) by taking into account that for
fixed $y$ one has $\lim_{a\rightarrow \infty }w/a=1$. At small distances
from the string, $mr\ll 1$, the leading term in the expansion of (\ref{TllM}%
) does not depend on the mass and coincides with (\ref{TllMm0}). The effect
of the mass is essential at distances $mr\gg 1$. In that region the VEVs are
suppressed by the factor $e^{-2mr\sin (\pi /q)}$ for $q\geq 2$ and $e^{-2mr}$
for $1\leq q<2$.

The special case $q=1$ corresponds to a magnetic flux in background of AdS
spacetime. The VEV of the energy-momentum tensor induced by the magnetic
flux we will denote as $\langle T_{\mu }^{\mu }\rangle _{\mathrm{mf}%
}=\langle T_{\mu }^{\mu }\rangle _{\mathrm{cs}}|_{q=1}$. For this case 
\begin{equation}
H(1,\alpha _{0},x)=\sin (\pi \alpha _{0})\sinh (2\alpha _{0}x)\sinh \left(
2x\right) ,  \label{h1}
\end{equation}%
and the general formula (\ref{Tllcs}) is simplified to (no summation over $%
\mu $)%
\begin{equation}
\left\langle T_{\mu }^{\mu }\right\rangle _{\mathrm{mf}}=\frac{\sin (\pi
\alpha _{0})}{\sqrt{2}\pi ^{7/2}a^{5}}\int_{0}^{\infty }dx\sinh \left(
2\alpha _{0}x\right) \tanh (x){\mathcal{F}}_{ma}(1+2\rho ^{2}\cosh ^{2}x),
\label{Tllmf}
\end{equation}%
for $\mu =0,1,3,4$, and $\left\langle T_{2}^{2}\right\rangle _{\mathrm{mf}%
}=\partial _{r}(r\left\langle T_{0}^{0}\right\rangle _{\mathrm{mf}})$.

Now let us consider the behavior of the string induced VEV in the asymptotic
regions of the radial coordinate $r$. In the region where the proper
distance from the string is much smaller than the curvature radius, $%
r_{p}\ll a$, one has $\rho \ll 1$. For the arguments of the function ${%
\mathcal{F}}_{ma}(u)$ in (\ref{Tllcs}) we have $u_{k},u_{x}\rightarrow 1+$.
By using the asymptotic (\ref{Fas}), we conclude that near the string the
effect of the mass is weak and to the leading order the VEV coincides with
that for a massless field. In that region the influence of the background
gravitational field on the string induced effects is weak and the VEV $%
\left\langle T_{\mu }^{\mu }\right\rangle _{\mathrm{cs}}$ behaves as $%
(w/r)^{5}$. By taking into account that the VEV depends on the coordinates $%
r$ and $w$ through the ratio $r/w$, from here we obtain the asymptotic near
the horizon for fixed value of $r$: near the horizon, $w\gg r$, the VEV
behaves like $w^{5}$. In the opposite limit of large distances from the
string one has $r_{p}\ll a$ and $\rho \gg 1$. By taking into account (\ref%
{Fas2}), the leading term in the asymptotic expansion is given as (no
summation over $\mu $)%
\begin{equation}
\left\langle T_{\mu }^{\mu }\right\rangle _{\mathrm{cs}}\approx \frac{%
(ma+3/2)(ma+1/2)}{2^{2ma+3}\pi ^{2}a^{5}(r/w)^{2ma+5}}h_{2ma+5}(q,\alpha
_{0})\ ,  \label{TllLarge}
\end{equation}%
for $\mu =0,1,3,4$ and $\langle T_{2}^{2}\rangle _{\mathrm{cs}}\approx
-2\left( ma+2\right) \langle T_{0}^{0}\rangle _{\mathrm{cs}}$. For a
massless field this result coincides with the exact formula (\ref{Tllm0}).
For a fixed $r$ and $w\ll r$, the expression (\ref{TllLarge}) describes the
behavior of the VEV\ near the AdS boundary: the VEV tends to zero as $%
w^{2ma+5}$. As seen from (\ref{TllLarge}), for a massive field on the AdS
bulk the decay of the string induced VEV at large distance from the string
follows a power law. This is in contrast to the case of the Minkowski bulk
where, as it has been mentioned above, the fall off is exponential.

\begin{figure}[tbph]
\begin{center}
\epsfig{figure=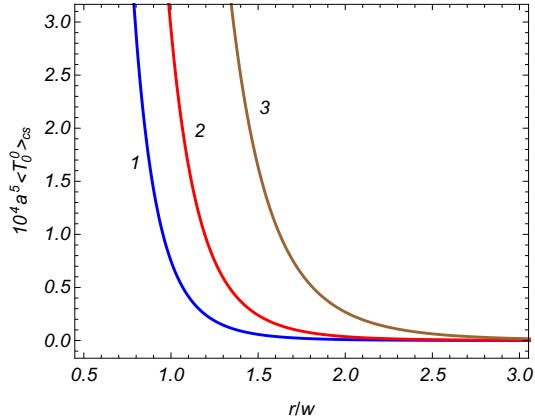,width=7.cm,height=5.5cm}
\end{center}
\caption{The dependence of the mean energy density, induced by the cosmic
string, on the radial coordinate for $ma=1$, $\protect\alpha _{0}=0.3$. The
numbers near the curves are the values of the parameter $q$. }
\label{fig1}
\end{figure}

In figure \ref{fig1} the energy density $\left\langle T_{0}^{0}\right\rangle
_{\mathrm{cs}}$ (in units of $1/a^{5}$) is plotted versus the proper
distance from the string (in units of the curvature scale $a$), $r_{p}/a=$ $%
r/w$. In the numerical evaluation we have taken $ma=1$ and $\alpha _{0}=0.3$%
. The corresponding values of the parameter $q$ are written near the curves.
In accordance with the asymptotic analysis given above, near the string the
energy density behaves as $(w/r)^{5}$ and at large distances the VEV is
suppressed by the factor $(w/r)^{2ma+5}$.

The dependence of the energy density induced by the string on the parameter $%
\alpha _{0}$ and on the mass (in units of $1/a$) is displayed in figure \ref%
{fig2} for $r/w=1.5$. The left panel is plotted for $ma=0.5$ and for the
right panel $\alpha _{0}=0.3$. For both panels, the numbers near the curves
present the corresponding values of $q$. As seen, for fixed values of the
other parameters, the absolute value of the energy density increases with
increasing planar angle deficit.

\begin{figure}[tbph]
\begin{center}
\begin{tabular}{cc}
\epsfig{figure=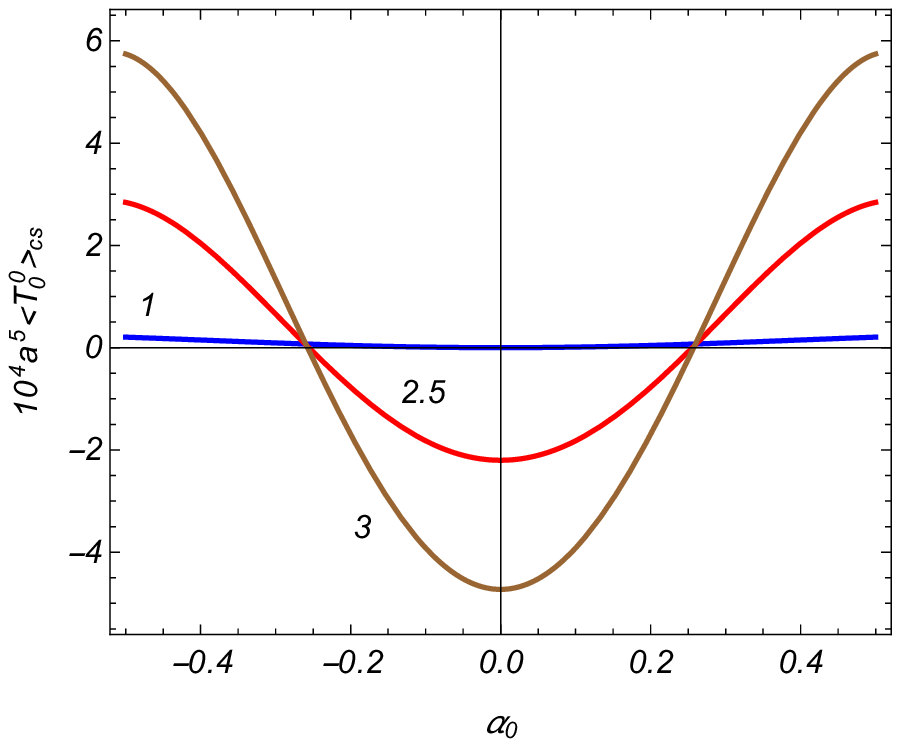,width=7.cm,height=5.5cm} & \quad %
\epsfig{figure=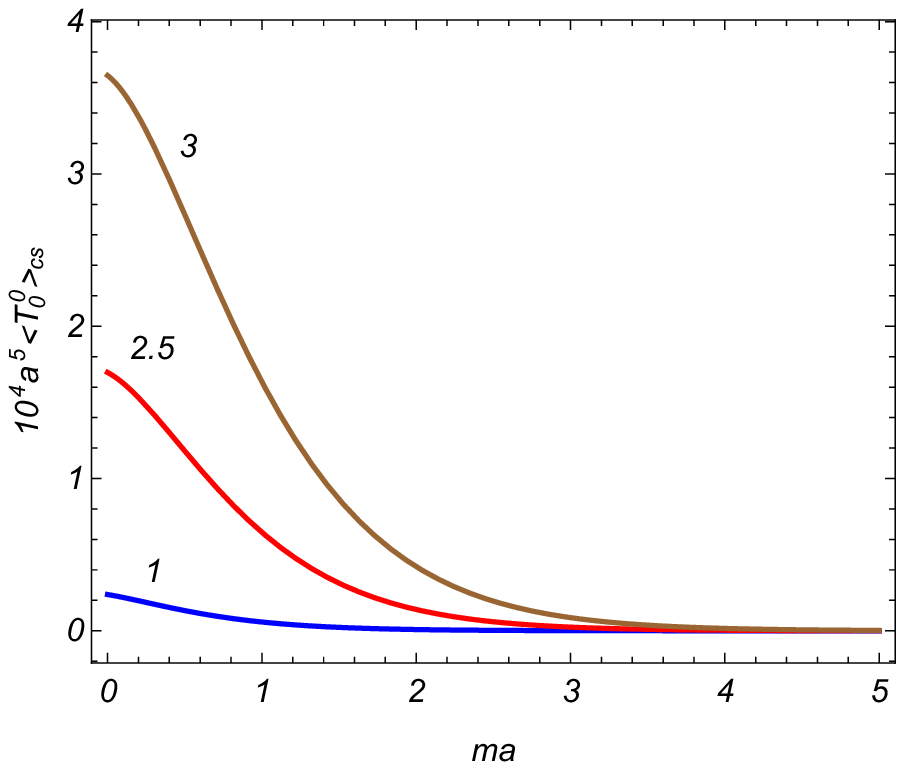,width=7.cm,height=5.5cm}%
\end{tabular}%
\end{center}
\caption{The energy density $\left\langle T_{0}^{0}\right\rangle _{\mathrm{cs%
}}$ as a function of the parameter $\protect\alpha _{0}$ (left panel) and of
the mass (right panel) for $r/w=1.5$. On the left panel $ma=0.5$ and on the
right one $\protect\alpha _{0}=0.3$. The numbers near the curves are the
values of $q$.}
\label{fig2}
\end{figure}

It is well known that in odd-dimensional spacetimes there are two
inequivalent irreducible representations of the Clifford algebra. Let us
denote the corresponding sets of Dirac gamma matrices as $\gamma _{(s)}^{(b)}
$ with $s=+1$ and $s=-1$ for separate representations. In the geometry under
consideration with (4+1)-dimensional spacetime we will take $\gamma
_{(+1)}^{(b)}=\gamma ^{(b)}$, where the matrices $\gamma ^{(b)}$ are given
by (\ref{Dirac04}), and for the second representation $\gamma
_{(-1)}^{(b)}=(-1)^{\delta _{4b}}\gamma ^{(b)}$. The Lagrangian density for
the field $\psi _{(s)}$ realizing the representation for a given $s$ reads $%
L_{(s)}=\bar{\psi}_{(s)}(i\gamma _{(s)}^{\mu }\mathcal{D}_{\mu }-m)\psi
_{(s)}$, where in the definition (\ref{DiracEq}) of the covariant derivative
the corresponding spin connection $\Gamma _{\mu }^{(s)}$ is taken. Note that 
$\Gamma _{\mu }^{(+1)}=\Gamma _{\mu }$. The operator of the energy-momentum
tensor has the form 
\begin{equation}
T_{\mu \nu }^{(s)}=\frac{i}{2}{\left[ \bar{\psi}_{(s)}\gamma _{(s)(\mu }%
\mathcal{D}_{\nu )}\psi _{(s)}-(\mathcal{D}_{(\mu }\bar{\psi}_{(s)})\gamma
_{(s)\nu )}\psi _{(s)}\right] }\ .  \label{EMTs}
\end{equation}%
Instead of the field $\psi _{(-1)}$ we can introduce a new field $\psi
^{\prime }=-\gamma ^{(4)}\psi _{(-1)}$. By taking into account that $\gamma
^{(4)}\gamma _{(-1)}^{\mu }\Gamma _{\mu }^{(-1)}\gamma ^{(4)}=\gamma ^{\mu
}\Gamma _{\mu }$, the corresponding Lagrangian density is rewritten in the
form $L_{(-1)}=\bar{\psi}^{\prime }\left( i\gamma ^{\mu }\mathcal{D}_{\mu
}+m\right) \psi ^{\prime }$ with $\mathcal{D}_{\mu }$ from (\ref{DiracEq}).
The equation of motion for the field $\psi ^{\prime }$ is given by (\ref%
{DiracEq}) with $s=-1$. The expression of the energy-momentum tensor $T_{\mu
\nu }^{(-1)}$, written in terms of the new field $\psi ^{\prime }$,
coincides with (\ref{EMTs}) for $s=+1$ replacing $\psi _{(+1)}$ by $\psi
^{\prime }$. As seen, the parameter $s$ introduced in (\ref{DiracEq}) is
interpreted in terms of the parameter that distinguishes inequivalent
representations. Hence, the VEV\ of the energy-momentum tensor $T_{\mu \nu
}^{(-1)}$ for the field $\psi _{(-1)}$ coincides with the VEV $\left\langle
T_{\mu \nu }\right\rangle $ investigated above for the value of the
parameter $s=-1$ in the Dirac equation (\ref{DiracEq}). From here we
conclude that the VEVs of the energy-momentum tensors coincide for fields
realizing two irreducible representations of the Clifford algebra. We recall
that the fermionic condensates for the fields $\psi _{(+1)}$ and $\psi
_{(-1)}$ have opposite signs \cite{Bell21FC}.

\section{Vacuum energy-momentum tensor induced by comapctification}

\label{sec:EMTcomp}

\subsection{General formula}

In this section we consider the VEV of the energy-momentum tensor in
background of the spacetime with the line element (\ref{ds1}) assuming that
the $z$-direction is compactified to a circle $S^{1}$ with the length $L$
and, hence, one has $0\leq z\leq L$. Introducing a new angular coordinate $%
\varphi $, $0\leq \varphi \leq 2\pi $, in accordance with $z=L\varphi /(2\pi
)$, the line element on the 3-dimensional hypersurface corresponding to the
string core $r=0$ is presented in the form%
\begin{equation}
ds_{\mathrm{core}}^{2}=e^{-2y/a}\left( dt^{2}-a^{2}\zeta ^{2}d\varphi
^{2}\right) -dy^{2}\ ,  \label{dsc}
\end{equation}%
where $\zeta =L/(2\pi a)$. The spatial line element is expressed as $dl_{%
\mathrm{core}}^{2}=dy^{2}+a^{2}\zeta ^{2}e^{-2y/a}d\varphi ^{2}$. A part of
the 2-dimensional spatial surface corresponding to this line element,
determined by $y_{0}\leq y<+\infty $, with $y_{0}=a\ln \zeta $, can be
embedded in a 3-dimensional Euclidean space with the coordinates $(X,Y,Z)$
by using the relations 
\begin{equation}
X=a\zeta e^{-y/a}\cos \varphi ,\;Y=a\zeta e^{-y/a}\sin \varphi
,\;Z=\int_{\ln \zeta }^{y/a}du\,\sqrt{1-\zeta ^{2}e^{-2u}}.  \label{XYZ}
\end{equation}%
In figure \ref{fig3} we have plotted the surface presenting the string core
with this embedding for $\zeta =1$. Note that the length of the compact
dimension measured by an observer having a fixed $y$-coordinate is given by $%
Le^{-y/a}$ and it exponentially decreases with increasing $y$.

\begin{figure}[tbph]
\begin{center}
\epsfig{figure=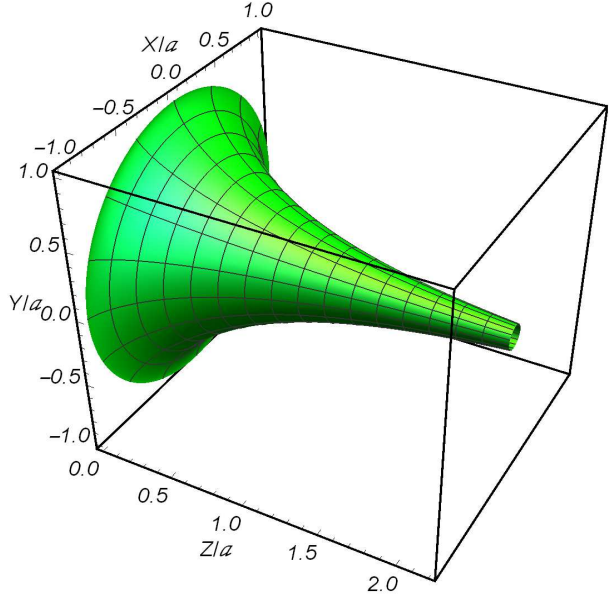,width=8.cm,height=8cm}
\end{center}
\caption{A part of the spatial surface corresponding to the string core $r=0$
with compactified $z$-coordinate, embedded in a 3-dimensional Euclidean
space. The surface is plotted for $\protect\zeta =1$. }
\label{fig3n}
\end{figure}

For the further discussion we need to specify the periodicity condition on
the field operator along the compac $z$-direction. Here, the quasi-periodic
condition 
\begin{equation}
\psi (t,r,\phi ,w,z+L)=e^{2\pi i\beta }\psi (t,r,\phi ,w,z),  \label{PCz}
\end{equation}%
will be imposed, where $\beta $ is an arbitrary constant. This leads to the
quantization of the eigenvalues for the momentum $k_{z}$: 
\begin{equation}
k_{z}=\tilde{k}_{l}=2\pi \frac{l+\tilde{\beta}}{L},\;l=0,\pm 1,\pm 2,\ldots ,
\label{kl}
\end{equation}%
where $\tilde{\beta}=\beta -\Phi _{\mathrm{c}}/\Phi _{0}$ and $\Phi _{%
\mathrm{c}}=-LA_{4}$. The component $A_{4}$ can be formally interpreted in
terms of the magnetic flux $\Phi _{\mathrm{c}}$ enclosed by the compact $z$%
-dimension.

The fermionic modes in the problem at hand are obtained from (\ref{Modes1})
taking $k_{z}=\tilde{k}_{l}$ and adding an additional factor $2\pi /L$ in
the expression (\ref{coeff}) for $|C_{\sigma }^{(\pm )}|^{2}$ (see also \cite%
{Bell20}). For the energy one has $E=\sqrt{\lambda ^{2}+p^{2}+\tilde{k}%
_{l}^{2}}$. The VEV of the energy-momentum tensor is evaluated by using the
formula (\ref{EMT}), where now the collective summation is understood as (%
\ref{sumj}) with the replacement $\int_{-\infty }^{\infty }dk_{z}\rightarrow
\sum_{l=-\infty }^{\infty }$. As before, the off-diagonal components vanish.
For the VEV of the diagonal components we find the representation (no
summation over $\mu $)%
\begin{equation}
\left\langle T_{\mu }^{\mu }\right\rangle =-\frac{qw^{6}}{2\pi a^{5}L}%
\sum_{j}\int_{0}^{\infty }d\lambda \lambda \int_{0}^{\infty
}dpp\sum_{l=-\infty }^{\infty }\left( \lambda ^{2}+p^{2}+\tilde{k}%
_{l}^{2}\right) ^{\delta _{0\mu }-1/2}(\tilde{k}_{l}^{2})^{\delta _{4\mu
}}R_{\beta _{j}}^{(\mu )}\left( \lambda r\right) W_{\nu }^{(\mu )}(pw).
\label{Tkkc}
\end{equation}%
The VEVs (\ref{Tkkc}) coincide for $s=+1$ and $s=-1$ and the consideration
below will be continued for $s=+1$. Here we are interested in the effects of
compactification and the corresponding contribution in the VEV can be
explicitly extracted by using the Abel-Plana formula \cite{Bell10AP}%
\begin{equation}
\sum_{l=-\infty }^{\infty }g(\tilde{k}_{l})=\frac{L}{\pi }\int_{0}^{\infty
}dx\,g\left( x\right) +\frac{iL}{2\pi }\int_{0}^{\infty }dx\sum_{n=\pm 1}%
\frac{g\left( ix\right) -g\left( -ix\right) }{e^{Lx+2\pi in\tilde{\beta}}-1}.
\label{AP1}
\end{equation}%
The part in the VEV $\left\langle T_{\mu }^{\mu }\right\rangle $ coming from
the first term in the right-hand side (\ref{AP1}) gives the corresponding
quantity in the uncompactified geometry and we get the representation (no
summation over $\mu $) 
\begin{equation}
\langle T_{\mu }^{\mu }\rangle =\langle T_{\mu }^{\mu }\rangle _{\mathrm{cs}%
}^{\mathrm{AdS}}+\langle T_{\mu }^{\mu }\rangle _{\mathrm{c}}\,.
\label{TmuDec}
\end{equation}%
The contribution in the VEV induced by compactification is expressed as 
\begin{eqnarray}
\left\langle T_{\mu }^{\mu }\right\rangle _{\mathrm{c}} &=&-\frac{%
(-1)^{\delta _{0\mu }+\delta _{4\mu }}qw^{6}}{2\pi ^{2}a^{5}}%
\sum_{j}\int_{0}^{\infty }d\lambda \lambda R_{\beta _{j}}^{(\mu )}\left(
\lambda r\right) \int_{0}^{\infty }dp\,pW_{\nu }^{(\mu )}(pw)  \notag \\
&&\times \int_{\sqrt{\lambda ^{2}+p^{2}}}^{\infty }dx\,x^{2\delta _{4\mu
}}\sum_{n=\pm 1}\frac{\left( x^{2}-\lambda ^{2}-p^{2}\right) ^{\delta _{0\mu
}-1/2}}{e^{Lx+2\pi in\tilde{\beta}}-1}.  \label{Tkkc2}
\end{eqnarray}%
The compactification does not change the local geometry and for $r>0$ the
renormalization in (\ref{TmuDec}) is only needed for the part $\langle
T_{\mu }^{\mu }\rangle _{\mathrm{cs}}^{\mathrm{AdS}}$. Consequently, the
implicit regularization, assumed in the discussion above, can be safely
removed in the topological part $\langle T_{\mu }^{\mu }\rangle _{\mathrm{c}%
} $.

By making use of the relation $(e^{u}-1)^{-1}=\sum_{l=1}^{\infty }e^{-lu}$
in (\ref{Tkkc2}), with $u=Lx+2\pi in\tilde{\beta}$, the integral over $x$ is
evaluated in terms of the modified Bessel function $K_{\nu }(lL\sqrt{\lambda
^{2}+p^{2}})$ with $\nu =0,1$. As the next step, we use the integral
representation 
\begin{equation}
K_{\nu }(z)=\frac{1}{2}\left( \frac{z}{2}\right) ^{\nu }\int_{0}^{\infty
}dt\,\frac{e^{-t-z^{2}/4t}}{t^{\nu +1}}\ .  \label{Macdonald}
\end{equation}%
The integrals over $\lambda $ and $p$ are evaluated in the way similar to
that we have discussed in the previous section for the part $\langle
T_{k}^{k}\rangle _{\mathrm{cs}}^{\mathrm{AdS}}$. Introducing the function (%
\ref{Jcal}), the following representation is obtained for the energy density%
\begin{equation}
\left\langle T_{0}^{0}\right\rangle _{\mathrm{c}}=\frac{q}{4\pi ^{2}a^{5}}%
\sum_{l=1}^{\infty }\cos (2\pi l\tilde{\beta})\int_{0}^{\infty
}dx\,x^{2}e^{-[1+\rho ^{2}+l^{2}L^{2}/(2w^{2})]x}[I_{\nu }(x)+I_{\nu +1}(x)]{%
\mathcal{J}}(q,\alpha _{0},\rho ^{2}x)\ .  \label{Tkkc3}
\end{equation}%
The remaining components are related to the energy density by the formulas%
\begin{equation}
\langle T_{1}^{1}\rangle _{\mathrm{c}}=\langle T_{3}^{3}\rangle _{\mathrm{c}%
}=\langle T_{0}^{0}\rangle _{\mathrm{c}},\;\langle T_{2}^{2}\rangle _{%
\mathrm{c}}=\left( 1+r\partial _{r}\right) \langle T_{0}^{0}\rangle _{%
\mathrm{c}},\;\left\langle T_{4}^{4}\right\rangle _{\mathrm{c}}=\partial
_{L}(L\left\langle T_{0}^{0}\right\rangle _{\mathrm{c}})\ .  \label{Tkkc4}
\end{equation}%
For the evaluation of the derivative $\partial _{u}{\mathcal{F}}_{ma}(u)$ in
the integrands for $\langle T_{2}^{2}\rangle _{\mathrm{c}}$ and $%
\left\langle T_{4}^{4}\right\rangle _{\mathrm{c}}$ we can use the relation (%
\ref{dF}).

Substituting the representation (\ref{Jf}) for function (\ref{Jcal}) in (\ref%
{Tkkc3}), the integral over $x$ is expressed in terms of the function (\ref%
{Ffunc}) and for the energy density we get%
\begin{eqnarray}
\left\langle T_{0}^{0}\right\rangle _{\mathrm{c}} &=&\frac{\sqrt{2}}{\pi
^{5/2}a^{5}}\sum_{l=1}^{\infty }\cos (2\pi l\tilde{\beta})\left[ %
\sideset{}{'_*}{\sum}_{k=0}^{[q/2]}(-1)^{k}c_{k}\cos (2\pi k\alpha _{0}){%
\mathcal{F}}_{ma}(u_{lk})\right.  \notag \\
&&+\left. \frac{q}{\pi }\int_{0}^{\infty }du\frac{H(q,\alpha _{0},u){%
\mathcal{F}}_{ma}(u_{lu})}{\cosh (2qu)-\cos (q\pi )}\right] \ ,
\label{T00c5}
\end{eqnarray}%
with the notations 
\begin{eqnarray}
&&u_{lk}=1+2\rho ^{2}s_{k}^{2}+\frac{l^{2}L^{2}}{2w^{2}}\ ,  \notag \\
&&u_{lu}=1+2\rho ^{2}\cosh ^{2}u+\frac{l^{2}L^{2}}{2w^{2}}\ .  \label{ulkx}
\end{eqnarray}%
The asterisk near the sign of summation over $k$ in (\ref{T00c5}) indicates
that the term $k=0$ must be divided by $2$. As a consequence of the maximal
symmetry of AdS spacetime the contribution $\left\langle T_{\mu }^{\mu
}\right\rangle _{\mathrm{c}}$ depends on $r$, $L$, $w$ through the
combinations $r/w$ and $L/w$. By using $\langle T_{1}^{1}\rangle _{\mathrm{c}%
}=\langle T_{3}^{3}\rangle _{\mathrm{c}}=\langle T_{0}^{0}\rangle _{\mathrm{c%
}}$, the covariant continuity equation $\nabla _{\mu }\left\langle T_{\nu
}^{\mu }\right\rangle _{\mathrm{c}}=0$ is reduced to the following two
relations $\langle T_{2}^{2}\rangle _{\mathrm{c}}=\partial _{r}(r\langle
T_{0}^{0}\rangle _{\mathrm{c}})$ and $\langle T_{2}^{2}\rangle _{\mathrm{c}%
}+\left\langle T_{4}^{4}\right\rangle _{\mathrm{c}}=(2-w\partial
_{w})\langle T_{0}^{0}\rangle _{\mathrm{c}}$. By taking into account that
the VEV $\left\langle T_{\mu }^{\mu }\right\rangle _{\mathrm{c}}$ is a
function of the ratios $r/w$ and $L/w$, it can be seen that these two
relations follow from the last two relations in (\ref{Tkkc4}). In addition,
we have the trace relation $\left\langle T_{\mu }^{\mu }\right\rangle _{%
\mathrm{c}}=sm\left\langle \bar{\psi}\psi \right\rangle _{\mathrm{c}}$.

We have considered two contributions in the VEV\ of the energy-momentum
tensor having a topological nature. The first one is generated by cosmic
string in the uncompactified geometry and the second one is induced by the
compactification. The combined topological part is expressed as $\langle
T_{\mu }^{\mu }\rangle _{\mathrm{t}}=\langle T_{\mu }^{\mu }\rangle _{%
\mathrm{cs}}+\langle T_{\mu }^{\mu }\rangle _{\mathrm{c}}$. Alternatively,
we can write $\langle T_{\mu }^{\mu }\rangle _{\mathrm{t}}=$ $\langle T_{\mu
}^{\mu }\rangle -\langle T_{\mu }^{\mu }\rangle _{\mathrm{AdS}}$. By using
the expressions for the separate parts, the topological contributions in the
components with $\mu =0,1,3$ are expressed as 
\begin{eqnarray}
\langle T_{\mu }^{\mu }\rangle _{\mathrm{t}} &=&\langle T_{\mu }^{\mu
}\rangle _{\mathrm{c}}^{(0)}+\frac{\sqrt{2}}{\pi ^{5/2}a^{5}}%
\sideset{}{'}{\sum}_{l=0}^{\infty }\cos (2\pi l\tilde{\beta})\left[ %
\sideset{}{'}{\sum}_{k=1}^{[q/2]}(-1)^{k}c_{k}\cos (2\pi k\alpha _{0}){%
\mathcal{F}}_{ma}(u_{lk})\right.  \notag \\
&&\left. +\frac{q}{\pi }\int_{0}^{\infty }du\frac{H(q,\alpha _{0},u){%
\mathcal{F}}_{ma}(u_{lu})}{\cosh (2qu)-\cos (q\pi )}\right] \ ,
\label{Tmut1}
\end{eqnarray}%
and the remaining components are determined from the relations $\langle
T_{2}^{2}\rangle _{\mathrm{t}}=\left( 1+r\partial _{r}\right) \langle
T_{0}^{0}\rangle _{\mathrm{t}}$ and$\;\left\langle T_{4}^{4}\right\rangle _{%
\mathrm{t}}=\partial _{L}(L\left\langle T_{0}^{0}\right\rangle _{\mathrm{t}%
}) $. The prime in $\sideset{}{'}{\sum}_{l=0}^{\infty }$ indicates that the
term with $l=0$ is taken with the coefficient 1/2. The first term in the
right-hand side of (\ref{Tmut1}) is defined as (no summation over $\mu $) 
\begin{equation}
\langle T_{\mu }^{\mu }\rangle _{\mathrm{c}}^{(0)}=\frac{a^{-5}}{\sqrt{2}\pi
^{5/2}}\sum_{l=1}^{\infty }\cos (2\pi l\tilde{\beta}){\mathcal{F}}%
_{ma}\left( 1+\frac{l^{2}L^{2}}{2w^{2}}\right) ,  \label{Tkk0}
\end{equation}%
for $\mu \neq 4$ and $\left\langle T_{4}^{4}\right\rangle _{\mathrm{c}%
}^{(0)}=\partial _{L}[L\left\langle T_{0}^{0}\right\rangle _{\mathrm{c}%
}^{(0)}]$. For $q=1$ and $\alpha _{0}=0$ the second term in the right-hand
side of (\ref{Tmut1}) vanishes and, hence, the part (\ref{Tkk0}) presents
the VEV in the geometry where the cosmic string is absent ((4+1)-dimensional
AdS spacetime with compactified $z$-direction). Consequently, the last term
in (\ref{Tmut1}) is induced by the planar angle deficit and by the magnetic
flux along the string's core. An alternative representation for the
topological part, given by (\ref{T00t2}), is provided in appendix \ref%
{sec:App1}.

\subsection{Special cases}

For a massless field one has the expression (\ref{Fm0}), and (\ref{T00c5})
is reduced to 
\begin{eqnarray}
\left\langle T_{0}^{0}\right\rangle _{\mathrm{c}} &=&\frac{6w^{5}}{\pi
^{2}a^{5}}\sum_{l=1}^{\infty }\cos (2\pi l\tilde{\beta})\left[ %
\sideset{}{'_*}{\sum}_{k=0}^{[q/2]}\frac{(-1)^{k}c_{k}\cos (2\pi k\alpha
_{0})}{(4r^{2}s_{k}^{2}+l^{2}L^{2})^{5/2}}\right.  \notag \\
&&+\left. \frac{q}{\pi }\int_{0}^{\infty }du\frac{H(q,\alpha _{0},u)}{\cosh
(2qu)-\cos (q\pi )}\frac{1}{(4r^{2}\cosh ^{2}u+l^{2}L^{2})^{5/2}}\right] \ .
\label{T00cm0}
\end{eqnarray}%
This result can be written as $\left\langle T_{0}^{0}\right\rangle _{\mathrm{%
c}}=(w/a)^{5}\left\langle T_{0}^{0}\right\rangle _{\mathrm{c}}^{\mathrm{(M)}%
} $, where $\left\langle T_{0}^{0}\right\rangle _{\mathrm{c}}^{\mathrm{(M)}}$
is the corresponding VEV in the Minkowski bulk. In the Minkowskian limit, $%
a\rightarrow \infty $, and for a massive field, by using (\ref{Funi}), we get%
\begin{eqnarray}
\left\langle T_{0}^{0}\right\rangle _{\mathrm{c}}^{\mathrm{(M)}} &=&\frac{%
2^{3/2}m^{5}}{\pi ^{5/2}}\sum_{l=1}^{\infty }\cos (2\pi l\tilde{\beta})\left[
\sideset{}{'_*}{\sum}_{k=0}^{[q/2]}(-1)^{k}c_{k}\cos (2\pi k\alpha
_{0})f_{5/2}(m\sqrt{4r^{2}s_{k}^{2}+l^{2}L^{2}})\right.  \notag \\
&&+\left. \frac{q}{\pi }\int_{0}^{\infty }du\,\frac{f_{5/2}(m\sqrt{%
4r^{2}\cosh ^{2}u+l^{2}L^{2}})}{\cosh (2qu)-\cos (q\pi )}H(q,\alpha _{0},u)%
\right] \ ,  \label{T00cM}
\end{eqnarray}%
where the function $f_{5/2}\left( x\right) $ is given by (\ref{f52}). For
large values of $L$ the dominant contribution in (\ref{T00cM}) comes from
the $l=1$ term and the VEV is exponentially suppressed by the factor $%
e^{-mL} $.

The case with $q=1$ and $\alpha _{0}=0$ corresponds to (4+1)-dimensional
locally AdS spacetime with a compactified $z$-coordinate in the absence of
cosmic string. In this special case the nonzero contribution to the
compactification part $\langle T_{\mu }^{\mu }\rangle _{\mathrm{c}}$ comes
from the $k=0$ term in (\ref{T00c5}). That term coincides with (\ref{Tkk0}).
The corresponding expression is further simplified for a massless field:%
\begin{equation}
\langle T_{\mu }^{\mu }\rangle _{\mathrm{c}}^{(0)}|_{m=0}=\frac{3\pi ^{-2}}{%
(aL/w)^{5}}\sum_{l=1}^{\infty }\frac{\cos (2\pi l\tilde{\beta})}{l^{5}}.
\label{Tkk0m0}
\end{equation}%
for $\mu \neq 4$ and $\left\langle T_{4}^{4}\right\rangle _{\mathrm{c}%
}^{(0)}=-4\left\langle T_{0}^{0}\right\rangle _{\mathrm{c}}^{(0)}$. This
expression is conformally related (with the conformal factor $(w/a)^{5}$) to
the corresponding result in (4+1)-dimensional Minkowski spacetime with a
single compact dimension of the length $L$. For small values of the proper
length of the compact dimension, compared with the curvature radius, one has 
$L/w\ll 1$ and by using the asymptotic expression (\ref{Fas}) we see that to
the leading order $\langle T_{\mu }^{\mu }\rangle _{\mathrm{c}}^{(0)}\approx
\langle T_{\mu }^{\mu }\rangle _{\mathrm{c}}^{(0)}|_{m=0}$. In this region
the effect of the mass is weak. In the opposite limit of large values of the
length of the compact dimension, $L/w\gg 1$, we use the approximation (\ref%
{Fas2}). In the leading order this gives%
\begin{equation}
\langle T_{\mu }^{\mu }\rangle _{\mathrm{c}}^{(0)}\approx \frac{%
(2ma+3)(2ma+1)}{\pi ^{2}a^{5}(L/w)^{2ma+5}}\sum_{l=1}^{\infty }\frac{\cos
(2\pi l\tilde{\beta})}{l^{2ma+5}},  \label{Tkk0L}
\end{equation}%
for $\mu \neq 4$ and $\left\langle T_{4}^{4}\right\rangle _{\mathrm{c}%
}^{(0)}\approx -2(ma+2)\left\langle T_{0}^{0}\right\rangle _{\mathrm{c}%
}^{(0)}$. For a massless field this result is exact.

Another special case corresponds to a magnetic flux in the absence of planar
angle deficit. In this case $q=1$ and the general formula (\ref{T00c5}) is
simplified to%
\begin{equation}
\left\langle T_{0}^{0}\right\rangle _{\mathrm{c}}=\left\langle
T_{0}^{0}\right\rangle _{\mathrm{c}}^{(0)}+\frac{\sqrt{2}\sin (\pi \alpha
_{0})}{\pi ^{7/2}a^{5}}\sum_{l=1}^{\infty }\cos (2\pi l\tilde{\beta}%
)\int_{0}^{\infty }du\,\sinh \left( 2\alpha _{0}u\right) \tanh (u){\mathcal{F%
}}_{ma}(u_{lu})\ ,  \label{T00q1}
\end{equation}%
with $\left\langle T_{0}^{0}\right\rangle _{\mathrm{c}}^{(0)}$ defined in (%
\ref{Tkk0}).

\subsection{Asymptotics and numerical results}

Now let us investigate the asymptotic behavior of the compactification
contribution at small and large distances from the string. For $2|\alpha
_{0}|<1-1/q$ we can directly put $r=0$ and one gets%
\begin{equation}
\left\langle T_{\mu }^{\mu }\right\rangle _{\mathrm{c}}|_{r=0}=\left[
1+2h_{0}(q,\alpha _{0})\right] \langle T_{\mu }^{\mu }\rangle _{\mathrm{c}%
}^{(0)},  \label{Tkkcr0}
\end{equation}%
for $\mu \neq 4$ and $\left\langle T_{4}^{4}\right\rangle _{\mathrm{c}%
}=[1+2h_{0}(q,\alpha _{0})]\partial _{L}(L\left\langle
T_{0}^{0}\right\rangle _{\mathrm{c}}^{(0)})$ at $r=0$. In this range of the
parameter $\alpha _{0}$ the part $\left\langle T_{\mu }^{\mu }\right\rangle
_{\mathrm{c}}$ is finite on the string. For $2|\alpha _{0}|>1-1/q$ the main
contribution in (\ref{T00c5}) comes from the integral term. In the integral
the contribution from large values of $u$ dominates. Introducing a new
integration variable $x=\rho ^{2}e^{2u}/2$, to the leading order we get 
\begin{eqnarray}
\left\langle T_{0}^{0}\right\rangle _{\mathrm{c}} &\approx &q\frac{%
2^{(|\alpha _{0}|-1/2)q-1}\cos [\pi q(1/2-|\alpha _{0}|)]}{\pi
^{7/2}a^{5}(r/w)^{1+(2|\alpha _{0}|-1)q}}\sum_{l=1}^{\infty }\cos (2\pi l%
\tilde{\beta})  \notag \\
&&\times \int_{0}^{\infty }dx\,x^{(|\alpha _{0}|-1/2)q-1/2}{\mathcal{F}}%
_{ma}\left( 1+x+\frac{l^{2}L^{2}}{2w^{2}}\right) \ .  \label{Tkkcr0b}
\end{eqnarray}%
In this case the compactification contributions in the VEVs diverge on the
string as $1/r^{1+(2|\alpha _{0}|-1)q}$.

At large distances from the string, corresponding to $r/w\gg 1$, to the
leading order we get $\left\langle T_{\mu }^{\mu }\right\rangle _{\mathrm{c}%
}\approx \langle T_{\mu }^{\mu }\rangle _{\mathrm{c}}^{(0)}$. The effects
induced by the planar angle deficit and by the magnetic flux appear in the
next order. In the corresponding expressions we use the asymptotic formula (%
\ref{Fas2}). In the expressions (\ref{ulkx}) we can ignore 1 and, then, the
series over $l$ is estimated as%
\begin{equation}
\sum_{l=1}^{\infty }\frac{\cos (2\pi l\tilde{\beta})}{\left(
b^{2}+l^{2}x^{2}\right) ^{ma+5/2}}\approx -\frac{1}{2b^{2ma+5}},
\label{Asser}
\end{equation}%
valid for $x\ll 1$. In this way we get%
\begin{equation}
\left\langle T_{0}^{0}\right\rangle _{\mathrm{c}}\approx \left\langle
T_{0}^{0}\right\rangle _{\mathrm{c}}^{(0)}-\frac{(2ma+3)(2ma+1)}{\pi
^{2}a^{5}(2r/w)^{2ma+5}}h_{2ma+5}(q,\alpha _{0})\ .  \label{Tkkrl}
\end{equation}%
Note that the correction induced by the planar angle deficit and by the
magnetic flux does not depend on the compactification radius. From here it
follows that $\left\langle T_{4}^{4}\right\rangle _{\mathrm{c}}-\left\langle
T_{4}^{4}\right\rangle _{\mathrm{c}}^{(0)}\approx \left\langle
T_{0}^{0}\right\rangle _{\mathrm{c}}-\left\langle T_{0}^{0}\right\rangle _{%
\mathrm{c}}^{(0)}$. The correction to the component $\langle
T_{2}^{2}\rangle _{\mathrm{c}}$ is obtained from (\ref{Tkkc4}). Combining
the asymptotic formulas (\ref{TllLarge}) and (\ref{Tkkrl}), we see that the
fall-off of the difference $\langle T_{\mu }^{\mu }\rangle _{\mathrm{t}%
}-\langle T_{\mu }^{\mu }\rangle _{\mathrm{c}}^{(0)}$ at large distances
from the string is stronger than the decay of separate terms $\left\langle
T_{\mu }^{\mu }\right\rangle _{\mathrm{cs}}$ and $\left\langle T_{\mu }^{\mu
}\right\rangle _{\mathrm{c}}-\left\langle T_{\mu }^{\mu }\right\rangle _{%
\mathrm{c}}^{(0)}$ in the topological part (\ref{Tmut1}) of the
energy-momentum tensor.

Now we consider the asymptotics with respect to the length of the compact
dimension $L$. For large values, $L\gg r,w$, the function ${\mathcal{F}}%
_{ma}(u)$ in the expression (\ref{T00c5}) for the compactification
contribution is approximated by (\ref{Fas2}). In the range $2|\alpha
_{0}|<1-1/q$ of the parameter $\alpha _{0}$, in (\ref{T00c5}) we can use the
approximations $u_{lk}\approx u_{lu}\approx (lL/w)^{2}/2$ and the leading
term reads 
\begin{equation}
\left\langle T_{\mu }^{\mu }\right\rangle _{\mathrm{c}}\approx \left[
1+2h_{0}(q,\alpha _{0})\right] \langle T_{\mu }^{\mu }\rangle _{\mathrm{c}%
}^{(0)},  \label{TmuclL}
\end{equation}%
for $\mu =0,1,2,3$, and $\left\langle T_{4}^{4}\right\rangle _{\mathrm{c}%
}\approx -2\left( ma+2\right) \left\langle T_{0}^{0}\right\rangle _{\mathrm{c%
}}$. The asymptotic for $\langle T_{\mu }^{\mu }\rangle _{\mathrm{c}}^{(0)}$
in (\ref{TmuclL}) is given by (\ref{Tkk0L}) and, to the leading order, the
effects of compactification do not depend on $r$. For the region $2|\alpha
_{0}|>1-1/q$, the contribution of the integral term in (\ref{T00c5})
dominates. The main contribution to the integral comes from the integration
range with $u\sim \ln (L/r)$. Replacing the integrand by its large $u$
asymptotic, one can show that $\langle T_{\mu }^{\mu }\rangle _{\mathrm{c}%
}\propto (w/L)^{2ma+5}(L/r)^{2|\alpha _{0}|q+1-q}$. Hence, for $2|\alpha
_{0}|>1-1/q$ the leading term in the compactification contribution depends
on the radial coordinate and its decay is weaker compared to the case $%
2|\alpha _{0}|<1-1/q$. As seen, the fall off of the contribution $%
\left\langle T_{\mu }^{\mu }\right\rangle _{\mathrm{c}}$ for large values of 
$L$ follows power law as a function of the length of the compact dimension.
Recall that for a massive field the corresponding decay in the Minkowski
bulk is exponential, like $e^{-mL}$.

For the investigation of the asymptotic in the region $L\ll w$, we use an
alternative representation (\ref{T00t2}) of the topological contribution.
The leading term of the part $\langle T_{0}^{0}\rangle _{\mathrm{c}}^{(0)}$
is given by the right-hand side of (\ref{Tkk0m0}) and it behaves as $\langle
T_{0}^{0}\rangle _{\mathrm{c}}^{(0)}\sim (w/L)^{5}$. The asymptotic behavior
of the difference $\langle T_{\mu }^{\mu }\rangle _{\mathrm{t}}-\langle
T_{\mu }^{\mu }\rangle _{\mathrm{c}}^{(0)}$ for small $L$ crucially depends
on the value of $\tilde{\beta}$. We have a periodicity with respect to that
parameter, with the period 1, and without loss of generality it will be
assumed that $|\tilde{\beta}|<1/2$. For $\tilde{\beta}=0$ the last
contribution in (\ref{T00t2}) is dominated by the $l=0$ term and, after
evaluation of the integral over $x$, for $\mu =0,1,3$ we get%
\begin{equation}
\langle T_{\mu }^{\mu }\rangle _{\mathrm{t}}\approx \langle T_{\mu }^{\mu
}\rangle _{\mathrm{c}}^{(0)}+\frac{\pi ^{-2}w}{a^{5}L}\left[ %
\sideset{}{'}{\sum}_{k=1}^{[q/2]}(-1)^{k}c_{k}\cos (2\pi k\alpha _{0}){%
\mathcal{F}}_{ma}^{(2)}(u_{k})+\frac{q}{\pi }\int_{0}^{\infty }dx\frac{%
H(q,\alpha _{0},x){\mathcal{F}}_{ma}^{(2)}(u_{x})}{\cosh (2qx)-\cos (q\pi )}%
\right] \ ,  \label{TmusLn}
\end{equation}%
where%
\begin{equation}
{\mathcal{F}}_{ma}^{(2)}(u)=\frac{Q_{ma-1}^{2}(u)+Q_{ma}^{2}(u)}{u^{2}-1}.
\label{Fma2}
\end{equation}%
The second term in the right-hand side of (\ref{TmusLn}), multiplied by $%
aL/w $, coincides with the VEV induced by cosmic string in (3+1)-dimensional
AdS spacetime (the corresponding line element is obtained from (\ref{ds1})
omitting $-dz^{2}$). In the special case $\alpha _{0}=0$ the latter
coincides with the result given in \cite{Beze13AdS}. For $\tilde{\beta}=0$
one has $\langle T_{0}^{0}\rangle _{\mathrm{c}}^{(0)}\approx $ $3\pi
^{-2}\zeta (5)(aL/w)^{-5}$and this term is dominant in (\ref{TmusLn}).

For $L\ll w$ and $0<|\tilde{\beta}|<1/2$, assuming that $|\tilde{\beta}|$ is
not too small, the integral over $x$ in (\ref{T00t2}) is dominated by the
contribution from the region with $x\sim w^{2}\pi |\tilde{\beta}-l|/(Lr)\gg 1
$. By using the asymptotic of the modified Bessel function for large
arguments \cite{Abra}, we can see that the leading contribution to the last
term in (\ref{T00t2}) coincides with the corresponding result for a massless
field. The latter is given by the last term in (\ref{T00t2m0}). If in
addition $L\ll r$, the dominant contribution comes from the $l=0$ term.
Replacing the modified Bessel function by its large argument asymptotic, we
can see that the difference $\langle T_{\mu }^{\mu }\rangle _{\mathrm{t}%
}-\langle T_{\mu }^{\mu }\rangle _{\mathrm{c}}^{(0)}$ behaves as $%
(rL/w^{2})^{-5/2}$ $\exp [-4\pi |\tilde{\beta}|\sin (\pi /q)r/L]$ for $q\geq
2$ and like $(rL/w^{2})^{-5/2}e^{-4\pi |\tilde{\beta}|r/L}$ for $1\leq q<2$.
Hence, we conclude that for small $L$ the topological contribution $\langle
T_{\mu }^{\mu }\rangle _{\mathrm{t}}$ is dominated by the part $\langle
T_{0}^{0}\rangle _{\mathrm{c}}^{(0)}$ with the leading behavior given by the
right-hand side of (\ref{Tkk0m0}).

It remains to consider the behavior of the VEV near the boundary and horizon
of the AdS spacetime. Near the boundary we have $w\ll L,r$ and the arguments
of the function ${\mathcal{F}}_{ma}(u)$ in (\ref{T00c5}) are large. By using
the asymptotic (\ref{Fas2}), it can be seen that the compactification
contribution $\left\langle T_{\mu }^{\mu }\right\rangle _{\mathrm{c}}$
vanishes on the AdS boundary like $w^{2ma+5}$. Near the horizon the ratios $%
w/L$ and $w/r$ are large and we use the asymptotic expression (\ref{Fas}).
The effect of the mass is weak and the leading term coincides with the VEV\
for a massless field. Hence, near the horizon the compactification induced
part behaves as $w^{5}$.

\begin{figure}[tbph]
\begin{center}
\begin{tabular}{cc}
\epsfig{figure=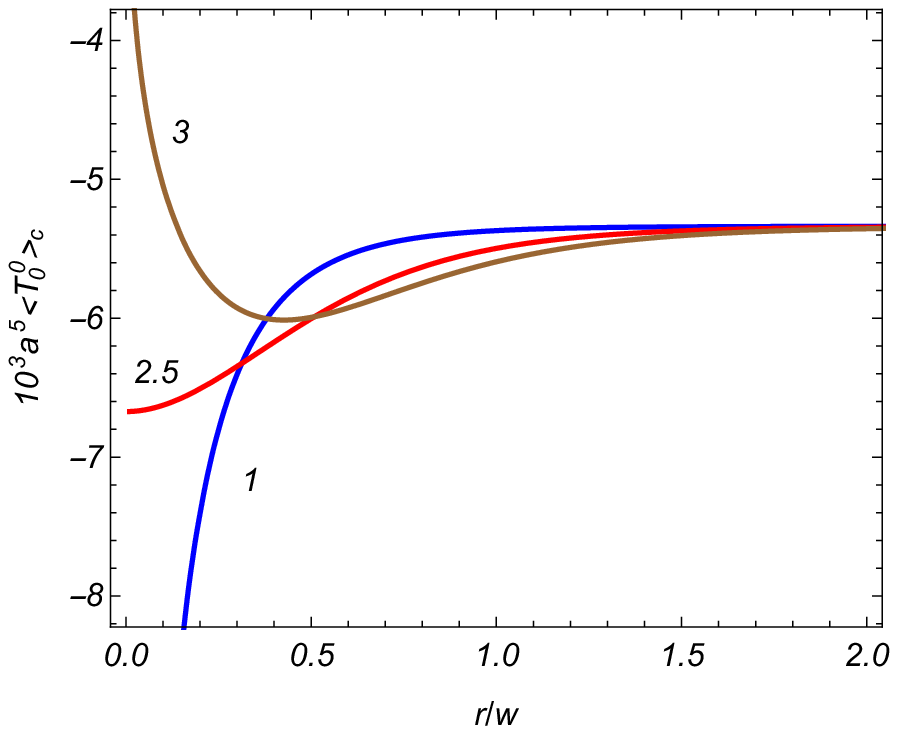,width=7.cm,height=5.5cm} & \quad %
\epsfig{figure=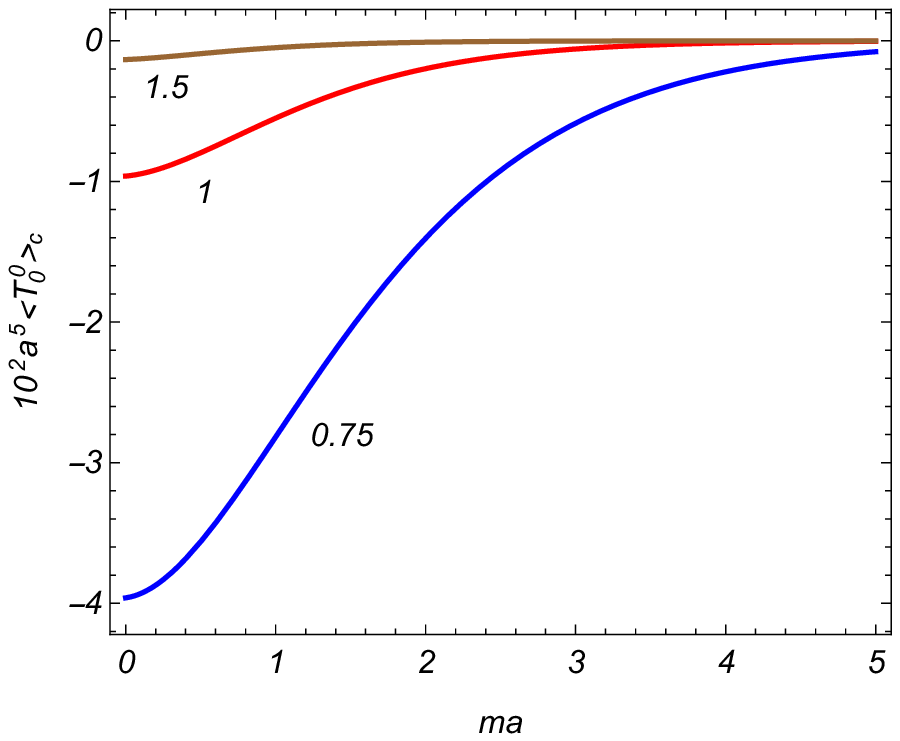,width=7.cm,height=5.5cm}%
\end{tabular}%
\end{center}
\caption{The contribution of the compactification in the VEV of the energy
density versus the radial coordinate (left panel) and the mass of the field
(right panel) for $\protect\alpha _{0}=0.3$, $\tilde{\protect\beta}=0.25$.
For the left panel $ma=1$, $L/w=1$ and the numbers near the curves present
the values of the parameter $q$. The right panel is plotted for $q=2.5$, $%
r/w=1$ and the numbers near the curves are the values of the ratio $L/w$.}
\label{fig3}
\end{figure}

Figure \ref{fig3} presents the dependence of the compactification part $%
\left\langle T_{0}^{0}\right\rangle _{\mathrm{c}}$ (in units of $1/a^{5}$)
on the radial coordinate (left panel) and on the mass of the field (right
panel). The left panel is plotted for $q=1,2.5,3$ (the numbers near the
curves) and for $\alpha _{0}=0.3$, $\tilde{\beta}=0.25$, $ma=1$, $L/w=1$. On
the right panel, the numbers near the curves correspond to the values of the
ratio $L/w$ (proper length of the compact dimension in units of the
curvature scale $a$) and the graphs are plotted for $\alpha _{0}=0.3$, $%
\tilde{\beta}=0.25$, $q=2.5$, $r/w=1$. For the graph on the left panel
corresponding to $q=1$ one has $2|\alpha _{0}|>1-1/q$ and, in accordance
with the asymptotic analysis given above, the VEV\ $\left\langle
T_{0}^{0}\right\rangle _{\mathrm{c}}$ diverges on the string as $%
1/(r/w)^{1+(2|\alpha _{0}|-1)q}$. For $q=2.5$ and $q=3$ the compactification
contribution is finite for $r=0$. The corresponding limiting values are
given by (\ref{Tkkcr0}) and they are equal to $-6.7\times 10^{-3}/a^{5}$ and 
$-2.4\times 10^{-3}/a^{5}$ for $q=2.5,3$, respectively. For $r/w\gg 1$ one
has $\left\langle T_{0}^{0}\right\rangle _{\mathrm{c}}-\left\langle
T_{0}^{0}\right\rangle _{\mathrm{c}}^{(0)}\propto (w/r)^{2ma+5}$ (see (\ref%
{Tkkrl})) and $\left\langle T_{0}^{0}\right\rangle _{\mathrm{c}}$ tends to $%
\left\langle T_{0}^{0}\right\rangle _{\mathrm{c}}^{(0)}$. The latter does
not depend on the parameters $q$ and $\alpha _{0}$ and in that region the
effects of the cosmic string are weak.

\begin{figure}[tbph]
\begin{center}
\begin{tabular}{cc}
\epsfig{figure=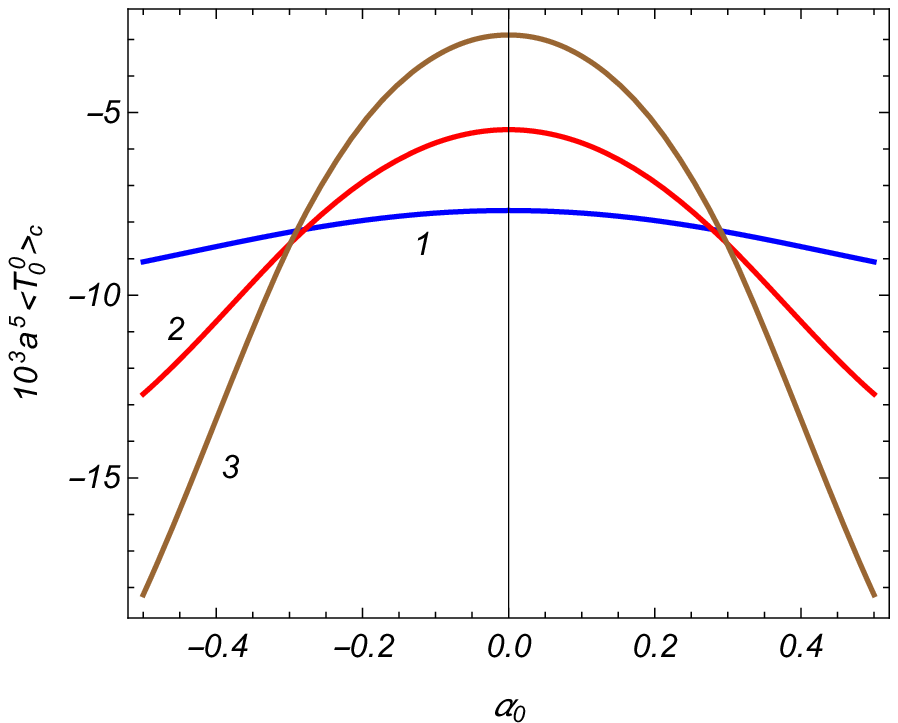,width=7.cm,height=5.5cm} & \quad %
\epsfig{figure=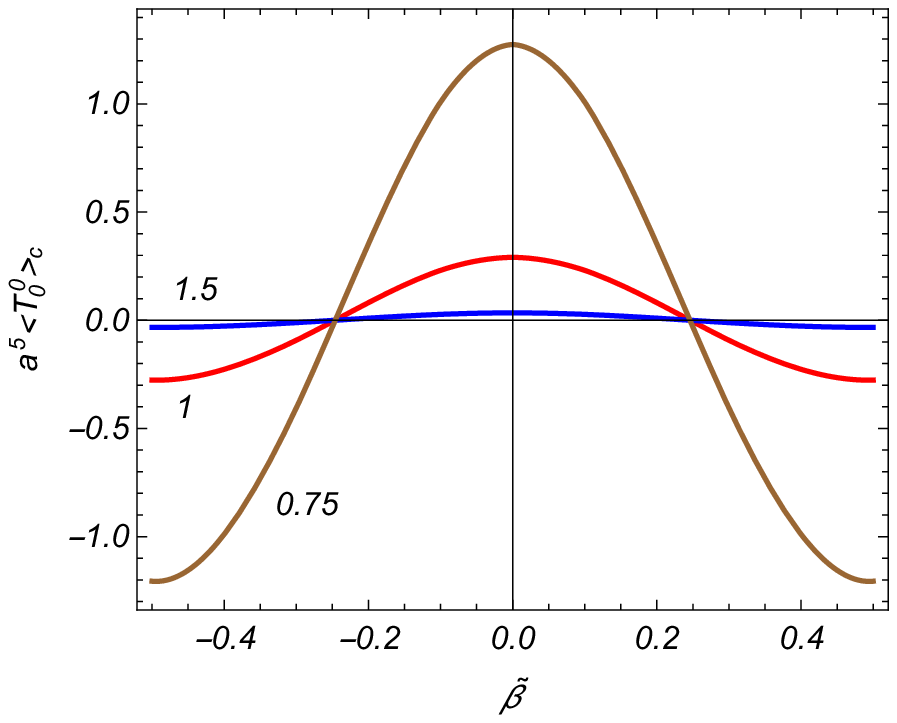,width=7.cm,height=5.5cm}%
\end{tabular}%
\end{center}
\caption{The VEV $\langle T_{0}^{0}\rangle _{\mathrm{c}}$ as a function of $%
\protect\alpha _{0}$ (left panel) and $\tilde{\protect\beta}$ (right panel)
fro $ma=0.5$, $r/w=0.5$. On the left panel, $\tilde{\protect\beta}=0.25$, $%
L/w=1$, and the numbers near the curves correspond to the values of $q$. For
the right panel $\protect\alpha _{0}=0.2$, $q=2$, and the numbers near the
curves are the values of $L/w$.}
\label{fig4}
\end{figure}

In figure \ref{fig4} we plot the VEV $\left\langle T_{0}^{0}\right\rangle _{%
\mathrm{c}}$ versus the parameters $\alpha _{0}$ (left panel) and $\tilde{%
\beta}$ (right panel). For the graphs on the left panel we have taken $%
\tilde{\beta}=0.25$, $ma=0.5$, $L/w=1$, $r/w=0.5$, and the numbers near the
curves represent the values of $q$. On the right panel $\alpha _{0}=0.2$, $%
q=2$, $ma=0.5$, $r/w=0.5$, and the numbers near the curves correspond to the
values of $L/w$. The compactification contribution $\left\langle T_{\mu
}^{\mu }\right\rangle _{\mathrm{c}}$ and its derivative with respect to the
parameter $\tilde{\beta}$ are continuous at points $\tilde{\beta}=\pm
1/2,\pm 3/2,\ldots $. As a function of the parameter $\alpha $, the VEV $%
\left\langle T_{\mu }^{\mu }\right\rangle _{\mathrm{c}}$ is continuous at
the points $\alpha =\pm 1/2,\pm 3/2,\ldots $, but the derivative $\partial
_{\alpha }\left\langle T_{\mu }^{\mu }\right\rangle _{\mathrm{c}}$is
discontinuous at those points.

\begin{figure}[tbph]
\begin{center}
\begin{tabular}{cc}
\epsfig{figure=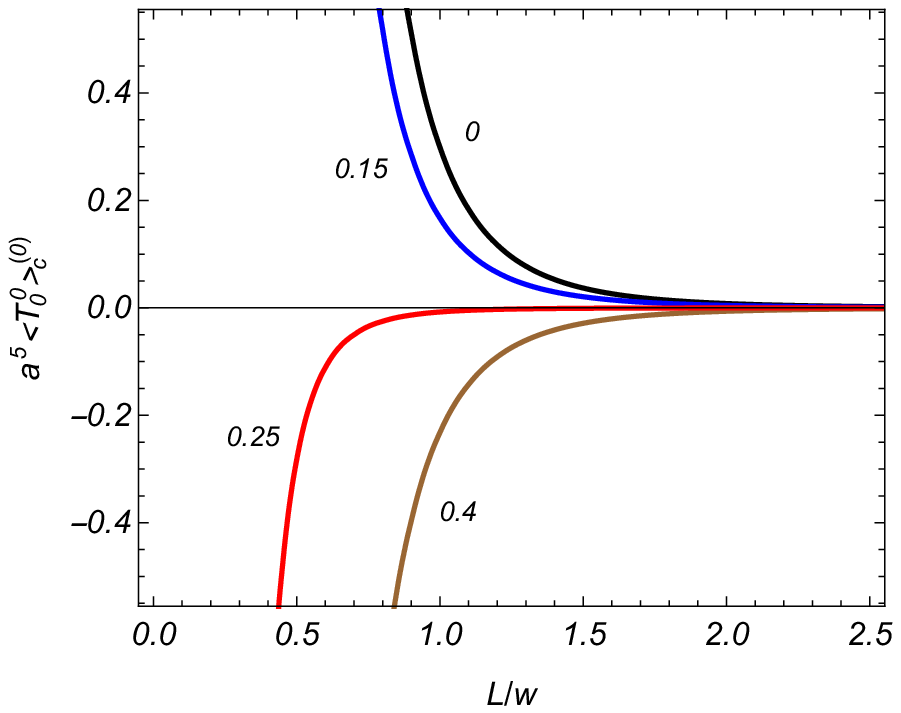,width=7.cm,height=5.5cm} & \quad %
\epsfig{figure=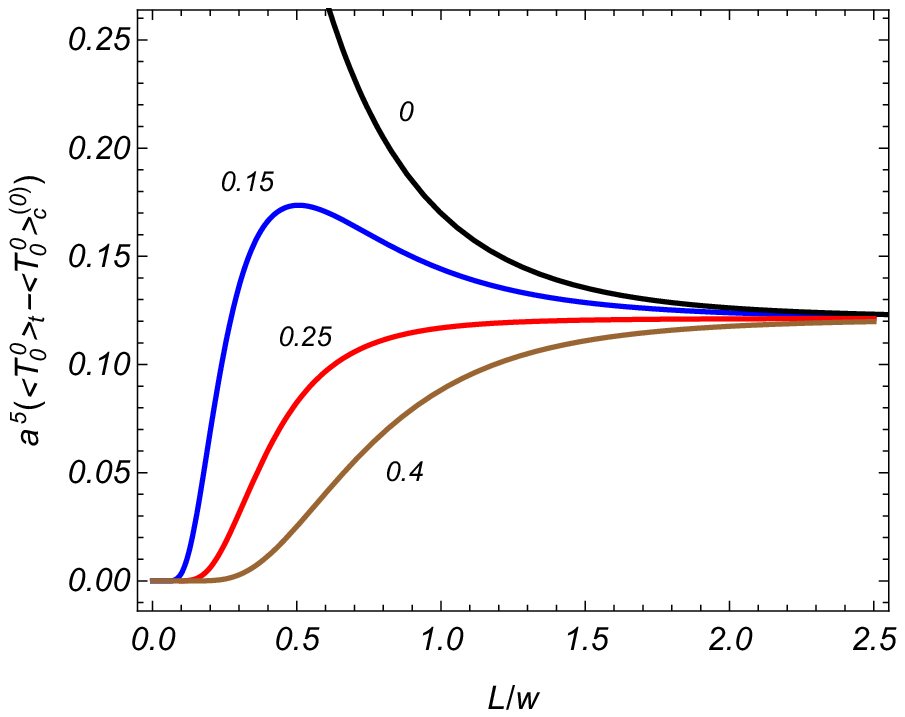,width=7.cm,height=5.5cm}%
\end{tabular}%
\end{center}
\caption{The topological part in the vacuum energy density in AdS spacetime
with a compact dimension, $\langle T_{0}^{0}\rangle _{\mathrm{c}}^{(0)}$,
(left panel) and the difference $\langle T_{0}^{0}\rangle _{\mathrm{t}%
}-\langle T_{0}^{0}\rangle _{\mathrm{c}}^{(0)}$ (right panel) versus the
proper length of the compact dimension (for the values of the parameters see
the text). }
\label{fig5}
\end{figure}

The dependence of the vacuum energy density on the proper length of the
compact dimensions (in units of the curvature radius $a$) is depicted in
figure \ref{fig5} for the values of the parameters $q=2.5$, $ma=0.5$, $%
\alpha _{0}=0.3$, $r/w=1$. The numbers near the curves correspond to the
values of $\tilde{\beta}$. On the left panel we have plotted the
contribution $\langle T_{0}^{0}\rangle _{\mathrm{c}}^{(0)}$ that corresponds
to the geometry in the absence of the cosmic string (locally AdS spacetime
with compactified $z$-dimension). The right panel presents the quantity $%
\langle T_{0}^{0}\rangle _{\mathrm{t}}-\langle T_{0}^{0}\rangle _{\mathrm{c}%
}^{(0)}$. It describes the effects induced in the topological part by the
cosmic string and magnetic flux along its axis. For small values of $L/w$
the effects of the mass are subdominant and the VEV on the left panel
behaves as $\langle T_{0}^{0}\rangle _{\mathrm{c}}^{(0)}\propto (L/w)^{-5}$.
In the opposite limit $L/w\gg 1$ one has $\langle T_{0}^{0}\rangle _{\mathrm{%
c}}^{(0)}\propto (w/L)^{2ma+5}$. As it has been explained by the asymptotic
analysis above, the behavior of the difference $\langle T_{0}^{0}\rangle _{%
\mathrm{t}}-\langle T_{0}^{0}\rangle _{\mathrm{c}}^{(0)}$ for small values
of $L/w$ is essentially different for $\tilde{\beta}=0$ and $0<|\tilde{\beta}%
|<1/2$. For $\tilde{\beta}=0$ that difference behaves like $w/L$, whereas
for $0<|\tilde{\beta}|<1/2$ it is exponentially suppressed (under the
condition $L\ll r$). For large values of the ratio $L/w$ the effects of
compactification are weak and $\langle T_{0}^{0}\rangle _{\mathrm{t}%
}-\langle T_{0}^{0}\rangle _{\mathrm{c}}^{(0)}$ tends to $\left\langle
T_{0}^{0}\right\rangle _{\mathrm{cs}}$. 

\section{Conclusion}

\label{sec:Conc}

We have studied the the influence of two sources of nontrivial topology on
the VEV of the energy-momentum tensor for a massive fermionic field
propagating in background of (4+1)-dimensional locally AdS spacetime. The
effects are induced by cosmic string and by the compactification of a
spatial dimension. An additional polarization of the vacuum is induced by
the background gravitational field. In odd number of spacetime dimensions
one has two inequivalent irreducible representations for the Clifford
algebra. We have shown that if the masses of the fields realizing different
representations are the same then the corresponding VEVs of the
energy-momentum tensor coincide. Note that the fermion condensates have
opposite signs for those fields.

We have divided the investigation in two stages. The first corresponds to
the geometry described by the line element (\ref{ds1}) with $-\infty
<z<+\infty $. The mean energy-momentum tensor is decomposed into two
contributions (see (\ref{Tmu1})). The first one is a purely gravitational
contribution and corresponds to the VEV in AdS spacetime. The latter
geometry is maximally symmetric and the part $\left\langle T_{\mu \nu
}\right\rangle ^{\mathrm{AdS}}$ is proportional to the metric tensor. The
second contribution, denoted here as $\left\langle T_{\mu \nu }\right\rangle
_{\mathrm{cs}}$, is induced by the cosmic string. For $r>0$ the local
geometry is not changed by the presence of the string and the
renormalization is needed only for the part $\left\langle T_{\mu \nu
}\right\rangle ^{\mathrm{AdS}}$. The vacuum energy-momentum tensor is
diagonal. The components $\left\langle T_{\mu }^{\mu }\right\rangle _{%
\mathrm{cs}}$ with $\mu =0,1,3,4$, given by (\ref{Tllcs}), coincide and the
component $\left\langle T_{2}^{2}\right\rangle _{\mathrm{cs}}$ is obtained
from the relation (\ref{T22cs}). Note that, unlike in the fermionic case,
for a massive scalar field the vacuum energy-momentum tensor will have a
nonzero off-diagonal component. As a consequence of the maximal symmetry of
the AdS spacetime, the string induced VEVs depend on the coordinates $r$ and 
$z$ in the form of the ratio $r/z$. The latter presents the proper distance
from the string measured in units of the curvature radius $a$. At small
distances from the string the effects of mass and of the background
gravitational field are weak and the leading term in the asymptotic
expansion is given by (\ref{Tllm0}) for the components $\mu =0,1,3,4$ and $%
\langle T_{2}^{2}\rangle _{\mathrm{cs}}\approx -4\langle T_{0}^{0}\rangle _{%
\mathrm{cs}}$. In that region the total VEV of the energy-momentum tensor is
dominated by the string induced contribution and behaves as $(w/r)^{5}$. At
large distances the influence of the gravitational field is essential and
the VEV $\left\langle T_{\mu }^{\mu }\right\rangle _{\mathrm{cs}}$ decays
like $(w/r)^{2ma+5}$. As a limiting case, we have obtained the VEV of the
energy-momentum tensor for a fermionic field, induced by a cosmic string in
background of (4+1)-dimensional Minkowski spacetime. It is expressed as (\ref%
{TllM}) with an exponential decay at large distances from the cosmic string.
The latter is in contrast to the power law fall off for the AdS bulk.
Another special case corresponds to the absence of planar angle deficit. In
this case the VEV is entirely induced by magnetic flux and is expressed as (%
\ref{Tllmf}).

The compactification of the coordinate $z$ on a circle with the length $L$
induces additional topological contributions to the characteristics of the
vacuum. The VEV of the corresponding energy density is given by the formula (%
\ref{T00c5}) and the other components are obtained by using the relations (%
\ref{Tkkc4}). An alternative representation of the topological part is
derived in Appendix \ref{sec:App1}. Note that in the compactified geometry
the components $\left\langle T_{0}^{0}\right\rangle _{\mathrm{c}}$ and $%
\left\langle T_{4}^{4}\right\rangle _{\mathrm{c}}$ differ. For a massless
field, we have the conformal relation $\left\langle T_{\nu }^{\mu
}\right\rangle _{\mathrm{c}}=(w/a)^{5}\left\langle T_{\nu }^{\mu
}\right\rangle _{\mathrm{c}}^{\mathrm{(M)}}$ with the corresponding VEV for
(4+1)-dimensional Minkowski bulk having spatial topology $R^{3}\times S^{1}$%
. In the case of a massive field the VEVs in the Minkwoski spacetime are
obtained taking the limit $a\rightarrow \infty $, with fixed value of the
coordinate $y$, and the corresponding energy density is expressed as (\ref%
{T00cM}). For a magnetic flux in the absence of planar angle deficit the
contribution in the energy density induced by compactification is given by (%
\ref{T00q1}). The compactification contribution in the VEV of the
energy-momentum tensor is finite on the cosmic string in the range of
parameters $2|\alpha _{0}|<1-1/q$ and diverges like $1/r^{1+(2|\alpha
_{0}|-1)q}$ for $2|\alpha _{0}|<1-1/q$. In both cases, near the string the
total VEV is dominated by the part $\left\langle T_{\mu }^{\mu
}\right\rangle _{\mathrm{cs}}$. At large distances from the cosmic string
the asymptotic of the compactification part in the VEV of the energy density
is described by (\ref{Tkkrl}). The last term, with the power law decay $%
(w/r)^{2ma+5}$, corresponds to the correction induced by the planar angle
deficit and by the magnetic flux and it does not depend on the
compactification radius. For large values of the compactification length and
for $2|\alpha _{0}|<1-1/q$ the leading term in the asymptotic expansion for $%
\left\langle T_{\mu }^{\mu }\right\rangle _{\mathrm{c}}$ is given by (\ref%
{TmuclL}) and the effects of compactification do not depend on $r$. In the
region $2|\alpha _{0}|>1-1/q$ the leading term in the compactification
contribution depends on the radial coordinate and its decay is weaker. For
small $L$ the leading term in the asymptotic expansion of the topological
contribution $\langle T_{\mu }^{\mu }\rangle _{\mathrm{t}}$ does not depend
on the planar angle deficit and on the magnetic flux and behaves as $%
(w/L)^{5}$. For small $L$ the behavior of the difference $\langle T_{\mu
}^{\mu }\rangle _{\mathrm{t}}-\langle T_{\mu }^{\mu }\rangle _{\mathrm{c}%
}^{(0)}$ crucially depends on the parameter $\tilde{\beta}$. For $\tilde{%
\beta}=0$ it decays like $w/L$ (see (\ref{TmusLn})), whereas for $0<|\tilde{%
\beta}|<1/2$ one has an exponential fall off as a function of $r/L$. The
compactification contribution $\left\langle T_{\mu }^{\mu }\right\rangle _{%
\mathrm{c}}$ vanishes on the AdS boundary like $w^{2ma+5}$ and behaves near
the horizon as $w^{5}$.

\section*{Acknowledgments}

A.A.S. was supported by the grant No. 20RF-059 of the Committee of Science
of the Ministry of Education, Science, Culture and Sport RA. E.R.B.M. is
partially supported by CNPQ under Grant no. 301.783/2019-3. W.O.S. thanks
Coordena\c{c}\~{a}o de Aperfei\c{c}oamento de Pessoal de N\'{\i}vel Superior
(CAPES) for financial support.

\appendix

\section{Alternative representation of the topological part}

\label{sec:App1}

An alternative representation for the topological contribution (\ref{Tmut1})
is obtained by using the relation (see also \cite{Bell21FC})%
\begin{equation}
\sideset{}{'}{\sum}_{l=0}^{\infty }\cos (2\pi l\tilde{\beta})e^{-\frac{%
l^{2}L^{2}}{2w^{2}}x}=\frac{w\sqrt{\pi }}{L\sqrt{2x}}\sum_{l=-\infty
}^{\infty }e^{-\frac{2\pi ^{2}w^{2}}{L^{2}x}\left( \tilde{\beta}-l\right)
^{2}},  \label{Rel}
\end{equation}%
and the representation (\ref{Ffunc}) for the function ${\mathcal{F}}_{ma}(u)$%
. This relation is a direct consequence of the Poisson resummation formula.
By taking into account the representation (\ref{Ffunc}) and (\ref{Rel}) we
get 
\begin{equation}
\sideset{}{'}{\sum}_{l=0}^{\infty }\cos (2\pi l\tilde{\beta}){\mathcal{F}}%
_{ma}\left( b+\frac{l^{2}L^{2}}{2w^{2}}\right) =\frac{\pi w}{2L}%
\int_{0}^{\infty }dx\,x^{3/2}e^{-bx}\left[ I_{\nu }(x)+I_{\nu +1}(x)\right]
\sum_{l=-\infty }^{\infty }e^{-\frac{2\pi ^{2}w^{2}}{L^{2}x}\left( \tilde{%
\beta}-l\right) ^{2}}.  \label{Resum}
\end{equation}%
Using this result in (\ref{Tmut1}), for the components with $\mu =0,1,3$ one
finds%
\begin{eqnarray}
\langle T_{\mu }^{\mu }\rangle _{\mathrm{t}} &=&\langle T_{\mu }^{\mu
}\rangle _{\mathrm{c}}^{(0)}+\frac{\pi ^{-3/2}w}{\sqrt{2}a^{5}L}%
\int_{0}^{\infty }dx\,x^{3/2}e^{-x}\left[ I_{\nu }(x)+I_{\nu +1}(x)\right]
\sum_{l=-\infty }^{\infty }e^{-\frac{2\pi ^{2}w^{2}}{L^{2}x}\left( \tilde{%
\beta}-l\right) ^{2}}  \notag \\
&&\times \left[ \sideset{}{'}{\sum}_{k=1}^{[q/2]}(-1)^{k}c_{k}\cos (2\pi
k\alpha _{0})e^{-2x\rho ^{2}s_{k}^{2}}+\frac{q}{\pi }\int_{0}^{\infty }du%
\frac{H(q,\alpha _{0},u)e^{-2x\rho ^{2}\cosh ^{2}u}}{\cosh (2qu)-\cos (q\pi )%
}\right] \ .  \label{T00t2}
\end{eqnarray}%
The other components are obtained from the relations $\langle
T_{2}^{2}\rangle _{\mathrm{t}}=\left( 1+r\partial _{r}\right) \langle
T_{0}^{0}\rangle _{\mathrm{t}}$ and$\;\left\langle T_{4}^{4}\right\rangle _{%
\mathrm{t}}=\partial _{L}(L\left\langle T_{0}^{0}\right\rangle _{\mathrm{t}})
$.

The representation (\ref{T00t2}) of the topological contribution is further
simplified for a massless fermionic field. By taking into account that $\nu
=-1/2$ and $I_{\nu }(x)+I_{\nu +1}(x)=e^{x}\sqrt{2/\pi x}$, after evaluating
the integral over $x$, we get%
\begin{eqnarray}
\langle T_{0}^{0}\rangle _{\mathrm{t}} &=&\langle T_{0}^{0}\rangle _{\mathrm{%
c}}^{(0)}+\frac{2\left( w/a\right) ^{5}}{L^{3}r^{2}}\sum_{l=-\infty
}^{\infty }(\tilde{\beta}-l)^{2}\left[ \sideset{}{'}{\sum}%
_{k=1}^{[q/2]}(-1)^{k}\frac{c_{k}}{s_{k}^{2}}\cos (2\pi k\alpha
_{0})K_{2}\left( 4\pi |\tilde{\beta}-l|s_{k}r/L\right) \right.  \notag \\
&&+\left. \frac{q}{\pi }\int_{0}^{\infty }du\frac{H(q,\alpha _{0},u)\cosh
^{-2}u}{\cosh (2qu)-\cos (q\pi )}K_{2}\left( 4\pi \left( r/L\right) |\tilde{%
\beta}-l|\cosh u\right) \right] \ .  \label{T00t2m0}
\end{eqnarray}%
Note that convergence of the series in this representation is stronger than
in (\ref{T00cm0}).


\begin{thebibliography}{99}
\bibitem{Hind95} M. B. Hindmarsh and T. W. B. Kibble, Cosmic strings.\ Rept.
Prog. Phys. \textbf{58}, 477 (1995).

\bibitem{Vile00} A. Vilenkin and E. P. S. Shellard, \textit{Cosmic Strings
and Other Topological Defects}\ (Cambridge University Press, Cambridge,
England, 2000).

\bibitem{Durr02} R. Durrer, M. Kunz, and A. Melchiorri, Cosmic structure
formation with topological defects.\ Phys. Rept. \textbf{364}, 1 (2002).

\bibitem{Bran21} R. Brandenberger, B. Cyr, and H. Jiao, Intermediate mass
black hole seeds from cosmic string loops. arXiv:2103.14057.

\bibitem{Sake07} M. Sakellariadou, Cosmic Strings. In: W. G. Unruh, R. Sch%
\"{u}tzhold (eds), \textit{Quantum Analogues: From Phase Transitions to
Black Holes and Cosmology}. Lecture Notes in Physics, vol 718 (Springer,
Berlin, Heidelberg, 2007).

\bibitem{Cope10} E. J. Copeland and T. W. B. Kibble, Cosmic strings and
superstrings. Proc. R. Soc. A \textbf{466}, 623 (2010).

\bibitem{Cope11} E.J. Copeland, L. Pogosian, and T.Vachaspati, Seeking
string theory in the cosmos. Class. Quantum Grav. \textbf{28}, 204009 (2011).

\bibitem{Cher15} D.F. Chernoff, S.-H. Henry Tye, Inflation, string theory
and cosmic strings. Int. J. Mod. Phys. D \textbf{24}, 1530010 (2015).

\bibitem{Hell86} T. M. Helliwell and D. A. Konkowski, Vacuum fluctuations
outside cosmic strings. Phys. Rev. D \textbf{34}, 1918 (1986).

\bibitem{Line87} B. Linet, Quantum field theory in the space-time of a
cosmic string. Phys. Rev. D \textbf{35}, 536 (1987).

\bibitem{Frol87} V. P. Frolov and E. M. Serebriany, Vacuum polarization in
the gravitational field of a cosmic string. Phys. Rev. D \textbf{35}, 3779
(1987).

\bibitem{Dowk87} J. S. Dowker, Casimir effect around a cone. Phys. Rev. D 
\textbf{36}, 3095 (1987).

\bibitem{Dowk87b} J. S. Dowker, Vacuum averages for arbitrary spin around a
cosmic string. Phys. Rev. D \textbf{36}, 3742 (1987).

\bibitem{Shir92} K. Shiraishi and S. Hirenzaki, Quantum aspects of
self-interacting fields around cosmic strings. Class. Quantum Grav. \textbf{9%
}, 2277 (1992).

\bibitem{Guim94} M. E. X. Guimar\~{a}es and B. Linet, Scalar Green's
functions in an Euclidean space with a conical-type line singularity.
Commun. Math. Phys. \textbf{165}, 297 (1994).

\bibitem{Cogn94} G. Cognola, K. Kirsten, and L. Vanzo, Free and
self-interacting scalar fields in the presence of conical singularities.
Phys. Rev. D \textbf{49}, 1029 (1994).

\bibitem{More95} E. S. Moreira Jnr., Massive quantum fields in a conical
background. Nucl. Phys. B \textbf{451}, 365 (1995).

\bibitem{Iell97} D. Iellici, Massive scalar field near a cosmic string.
Class. Quantum Grav. \textbf{14}, 3287 (1997).

\bibitem{Khus99} N. R. Khusnutdinov and M. Bordag, Ground state energy of a
massive scalar field in the background of a cosmic string of finite
thickness. Phys. Rev. D \textbf{59}, 064017 (1999).

\bibitem{Beze06cyl} E. R. Bezerra de Mello, V. B. Bezerra, A. A. Saharian,
and A. S. Tarloyan, Vacuum polarization induced by a cylindrical boundary in
the cosmic string spacetime. Phys. Rev. D \textbf{74}, 025017 (2006).

\bibitem{Beze06} V. B. Bezerra and N. R. Khusnutdinov, The vacuum
expectation value of the spinor massive field in the cosmic string
spacetime. Class. Quantum Grav. \textbf{23}, 3449 (2006).

\bibitem{Beze08cyl} E. R. Bezerra de Mello, V. B. Bezerra, A. A. Saharian,
and A. S. Tarloyan, Fermionic vacuum polarization by a cylindrical boundary
in the cosmic string spacetime. Phys. Rev. D \textbf{78}, 105007 (2008).

\bibitem{Beze08compos} E. R. Bezerra de Mello and A. A. Saharian, Fermionic
vacuum polarization by a composite topological defect in higher-dimensional
space-time. Phys. Rev. D \textbf{78}, 045021 (2008).

\bibitem{Site12} Yu. A. Sitenko and N. D. Vlasii, Induced vacuum
energy-momentum tensor in the background of a cosmic string. Class. Quantum
Grav. 29, 095002 (2012).

\bibitem{Brev95} I. Brevik and T. Toverud, Electromagnetic energy density
around a superconducting cosmic string. Class. Quantum Grav. \textbf{12},
1229 (1995).

\bibitem{Beze07cyl} E. R. Bezerra de Mello, V. B. Bezerra, and A. A.
Saharian, Electromagnetic Casimir densities induced by a conducting
cylindrical shell in the cosmic string spacetime. Phys. Lett. B \textbf{645}%
, 245 (2007).

\bibitem{Nest11} V. V. Nesterenko and I. G. Pirozhenko, Vacuum energy in
conical space with additional boundary conditions. Class. Quantum Grav. 
\textbf{28}, 175020 (2011).

\bibitem{Fucc11} G. Fucci and K. Kirsten, The Casimir effect for conical
pistons. J. High Energy Phys. 03(2011)016.

\bibitem{Beze12ring} E. R. Bezerra de Mello, F. Moraes, and A. A. Saharian,
Fermionic Casimir densities in a conical space with a circular boundary and
magnetic flux. Phys. Rev. D \textbf{85}, 045016 (2012).

\bibitem{Beze11perp} E. R. Bezerra de Mello and A. A. Saharian, Vacuum
polarization by a flat boundary in cosmic string spacetime. Class. Quantum
Grav. \textbf{28}, 145008 (2011).

\bibitem{Beze12perp} E. R. Bezerra de Mello, A. A. Saharian, and A. Kh.
Grigoryan, Casimir effect for parallel metallic plates in cosmic string
spacetime. J. Phys. A \textbf{45}, 374011 (2012).

\bibitem{Beze13perp} E. R. Bezerra de Mello, A. A. Saharian, and S. V.
Abajyan, Fermionic vacuum polarization by a flat boundary in cosmic string
spacetime. Class. Quantum Grav. \textbf{30}, 015002 (2013).

\bibitem{Beze18perp} E. R. Bezerra de Mello, A. A. Saharian, and S. V.
Abajyan, Scalar Casimir densities and forces for parallel plates in cosmic
string spacetime. Phys. Rev. D \textbf{97}, 085023 (2018).

\bibitem{Beze12top} E. R. Bezerra de Mello and A. A. Saharian, Topological
Casimir effect in compactified cosmic string spacetime. Class. Quantum Grav. 
\textbf{29}, 035006 (2012).

\bibitem{Bell14top} S. Bellucci, E. R. Bezerra de Mello, A. de Padua, and A.
A. Saharian, Fermionic vacuum polarization in compactified cosmic string
spacetime. Eur. Phys. J. C \textbf{74}, 2688 (2014).

\bibitem{Brag19} E. A. F. Bragan\c{c}a, H. F. Santana Mota, and E. R.
Bezerra de Mello, Vacuum expectation value of the energy-momentum tensor in
a higher-dimensional compactified cosmic string spacetime. Eur. Phys. J.
Plus \textbf{134}, 400 (2019).

\bibitem{Davi88} P. C. W. Davies and V. Sahni, Quantum gravitational effects
near cosmic strings. Class. Quantum Grav. \textbf{5}, 1 (1988).

\bibitem{Otte10} A. C. Ottewill and P. Taylor, Vacuum polarization on the
Schwarzschild metric threaded by a cosmic string. Phys. Rev. D \textbf{82},
104013 (2010).

\bibitem{Otte11} A. C. Ottewill and P. Taylor, Renormalized vacuum
polarization and stress tensor on the horizon of a Schwarzschild black hole
threaded by a cosmic string. Class. Quantum Grav. \textbf{28}, 015007 (2011).

\bibitem{Beze09dS} E. R. Bezerra de Mello and A. A. Saharian, Vacuum
polarization by a cosmic string in de Sitter spacetime. J. High Energy Phys.
04(2009)046.

\bibitem{Beze10dS} E. R. Bezerra de Mello and A. A. Saharian, Fermionic
vacuum polarization by a cosmic string in de Sitter spacetime. J. High
Energy Phys. 08(2010)038.

\bibitem{Saha17dS} A. A. Saharian, V. F. Manukyan, and N. A. Saharyan,
Electromagnetic vacuum fluctuations around a cosmic string in de Sitter
spacetime. Eur. Phys. J. C \textbf{77}, 478 (2017).

\bibitem{Brag20dS} E. A. F. Bragan\c{c}a, E.R. Bezerra de Mello, and A.
Mohammadi, Induced fermionic vacuum polarization in a de Sitter spacetime
with a compactified cosmic string. Phys. Rev. D \textbf{101}, 045019 (2020).

\bibitem{Beze12AdS} E. R. Bezerra de Mello and A. Saharian, Vacuum
polarization induced by a cosmic string in anti-de Sitter spacetime. J.
Phys. A \textbf{45}, 115402 (2012).

\bibitem{Beze13AdS} E. R. Bezerra de Mello, E. R. Figueiredo Medeiros, and
A. A. Saharian, Fermionic vacuum polarization by a cosmic string in anti-de
Sitter spacetime. Class. Quantum Grav. \textbf{30}, 175001 (2013).

\bibitem{Oliv20compAdS} W. Oliveira dos Santos, E. R. Bezerra de Mello, and
H. F. Mota, Vacuum polarization in high-dimensional AdS space-time in the
presence of a cosmic string and a compactified extra dimension. Eur. Phys.
J. Plus \textbf{135}, 27 (2020).

\bibitem{Bell20} S. Bellucci, W. Oliveira dos Santos, and E.R. Bezerra de
Mello, Induced fermionic current in AdS spacetime in the presence of a
cosmic string and a compactified dimension, Eur. Phys. J. C \textbf{80}, 963
(2020).

\bibitem{Bell21FC} S. Bellucci, W. Oliveira dos Santos, E. R. Bezerra de
Mello, and A. A. Saharian, Topological effects in fermion condensate induced
by cosmic string and compactification on AdS bulk. arXiv:2105.00829.

\bibitem{Bell21AdSb} S. Bellucci, W. Oliveira dos Santos, E. R. Bezerra de
Mello, and A. A. Saharian, Vacuum fermionic currents in braneworld models on
AdS bulk with a cosmic string. J. High Energy Phys. 02(2021)190.

\bibitem{Maar10} R. Maartens and K. Koyama, Brane-world gravity. Living Rev.
Relativity \textbf{13}, 5 (2010).

\bibitem{Ahar00} O. Aharony, S. S. Gubser, J. Maldacena, H. Ooguri, and Y.
Oz, Large N field theories, string theory and gravity. Phys. Rep. \textbf{323%
}, 183 (2000).

\bibitem{Ammo15} M. Ammon and J. Erdmenger, \textit{Gauge/Gravity Duality:
Foundations and Applications} (Cambridge University Press, Cambridge,
England, 2015).

\bibitem{Dehg02} M. H. Dehghani, A. M. Ghezelbash, and R. B. Mann, Vortex
holography. Nucl. Phys. B \textbf{625}, 389 (2002).

\bibitem{Ball11} C. A. Ballon Bayona, C. N. Ferreira, and V. J. Vasquez
Otoya, A conical deficit in the AdS$_{4}$/CFT$_{3}$ correspondence. Class.
Quantum Grav. \textbf{28}, 015011 (2011).

\bibitem{Padu16} A. de P\'{a}dua Santos and E.R. Bezerra de Mello,
Non-Abelian cosmic strings in de Sitter and anti-de Sitter space. Phys. Rev.
D \textbf{94}, 063524 (2016).

\bibitem{Abra} \textit{Handbook of Mathematical Functions}, edited by M.
Abramowitz and I. A. Stegun (Dover, New York, 1972).

\bibitem{Nist10} F.W. Olver et al, \textit{NIST Handbook of Mathematical
Functions} (Cambridge University Press, USA, 2010).

\bibitem{Beze10f3} E. R. Bezerra de Mello, V. B. Bezerra, A. A. Saharian,
and V. M. Bardeghyan, Fermionic current densities induced by magnetic flux
in a conical space with a circular boundary. Phys. Rev. D \textbf{82},
085033 (2010).

\bibitem{Bell10AP} S. Bellucci, A. Saharian, and V. Bardeghyan, Induced
fermionic current in toroidally compactified spacetimes with applications to
cylindrical and toroidal nanotubes. Phys. Rev. D \textbf{82}, 065011 (2010).
\end{thebibliography}
\end{document}